\documentclass[twocolumn,english,aps,prb,showpacs,byrevtex,amsmath,amssymb,superscriptaddress,longbibliography]{revtex4-1}
\usepackage[T1]{fontenc}
\usepackage[latin9]{inputenc}
\usepackage{lineno}
\usepackage{textcomp}
\usepackage{amsmath}
\usepackage{graphicx}
\usepackage{multirow}
\usepackage{ bbold }
\usepackage{lipsum}
\usepackage{array}
\usepackage{physics}
\newcolumntype{P}[1]{>{\centering\arraybackslash}p{#1}}
\usepackage[title]{appendix}

\makeatletter
\usepackage{lipsum}
\usepackage{epsfig}

\usepackage[colorlinks,linkcolor=blue]{hyperref}
\usepackage[nameinlink,capitalise]{cleveref}

\usepackage{subfigure}
\usepackage{color}
\usepackage{physics}

\definecolor{Red}{rgb}{1,0,0}
\definecolor{Blu}{rgb}{0,0,1}
\definecolor{Green}{rgb}{0,1,0}

\makeatother
\usepackage{amssymb}
\usepackage{babel}

\usepackage{tikz,xcolor,hyperref}
\definecolor{lime}{HTML}{A6CE39}
\DeclareRobustCommand{\orcidicon}{%
	\begin{tikzpicture}
	\draw[lime, fill=lime] (0,0)
	circle [radius=0.16]
	node[white] {{\fontfamily{qag}\selectfont \tiny ID}};
	\draw[white, fill=white] (-0.0625,0.095)
	circle [radius=0.007];
	\end{tikzpicture}
	\hspace{-2mm}
}

\foreach \x in {A, ..., Z}{%
	\expandafter\xdef\csname orcid\x\endcsname{\noexpand\href{https://orcid.org/\csname orcidauthor\x\endcsname}{\noexpand\orcidicon}}
}

\begin{document}

\title{Topological states in superlattices of HgTe class of materials 
\\ for engineering three-dimensional flat bands}


\author{Rajibul Islam\orcidC}
\email{rislam@magtop.ifpan.edu.pl}
\affiliation{International Research Centre MagTop, Institute of Physics, Polish Academy of Sciences, Aleja Lotnik\'ow 32/46, PL-02668 Warsaw, Poland}

\author{Barun Ghosh}
\affiliation{Department of Physics, Northeastern University, Boston, Massachusetts 02115, USA}
\affiliation{Department of Physics, Indian Institute of Technology, Kanpur 208016, India}

\author{Giuseppe Cuono\orcidD}
\affiliation{International Research Centre MagTop, Institute of Physics, Polish Academy of Sciences, Aleja Lotnik\'ow 32/46, PL-02668 Warsaw, Poland}

\author{Alexander Lau\orcidE}
\affiliation{International Research Centre MagTop, Institute of Physics, Polish Academy of Sciences, Aleja Lotnik\'ow 32/46, PL-02668 Warsaw, Poland}

\author{Wojciech Brzezicki\orcidG}
\affiliation{Institute of Theoretical Physics, Jagiellonian University, ulica S. \L{}ojasiewicza 11, PL-30348 Krak\'ow, Poland}
\affiliation{International Research Centre MagTop, Institute of Physics, Polish Academy of Sciences, Aleja Lotnik\'ow 32/46, PL-02668 Warsaw, Poland}

\author{Arun Bansil\orcidH}
\affiliation{Department of Physics, Northeastern University, Boston, Massachusetts 02115, USA}

\author{Amit Agarwal}
\affiliation{Department of Physics, Indian Institute of Technology, Kanpur 208016, India}

\author{Bahadur Singh\orcidF}
\email{bahadur.singh@tifr.res.in}
\affiliation{Department of Condensed Matter Physics and Materials Science, Tata Institute of Fundamental Research, Colaba, Mumbai 400005, India.}

\author{Tomasz Dietl\orcidB}
\affiliation{International Research Centre MagTop, Institute of Physics, Polish Academy of Sciences, Aleja Lotnik\'ow 32/46, PL-02668 Warsaw, Poland}
\affiliation{WPI-Advanced Institute for Materials Research, Tohoku University, Sendai 980-8577, Japan}

\author{Carmine Autieri\orcidA}
\email{autieri@magtop.ifpan.edu.pl}
\affiliation{International Research Centre MagTop, Institute of Physics, Polish Academy of Sciences, Aleja Lotnik\'ow 32/46, PL-02668 Warsaw, Poland}
\affiliation{Consiglio Nazionale delle Ricerche CNR-SPIN, UOS Salerno, IT-84084 Fisciano (Salerno),
Italy}

\begin{abstract}
In search of materials with three-dimensional flat band dispersions, using {\em ab-initio} computations we investigate how topological phases evolve as a function of hydrostatic pressure and uniaxial strain in two types of superlattices: HgTe/CdTe and HgTe/HgSe. In short-period HgTe/CdTe superlattices, our analysis unveils the presence of isoenergetic nodal lines, which could host strain-induced three-dimensional flat bands at the Fermi level without requiring doping, when fabricated,
for instance, as core-shell nanowires. In contrast, HgTe/HgSe short-period superlattices are found to harbor a rich phase diagram with a plethora of topological phases. Notably, the unstrained superlattice realizes an ideal Weyl semimetal with Weyl points situated at the Fermi level. A small-gap topological insulator with multiple band inversions can be obtained by tuning the volume: under compressive uniaxial strain, the material transitions sequentially into a Dirac semimetal to a nodal-line semimetal, and finally into a topological insulator with a single band inversion.
\end{abstract}
\date{\today}
\maketitle

\section{Introduction}
\label{sec:intro}

In the past decades, tremendous effort has been made both theoretically and experimentally to search, predict, and understand the characteristics of a wide variety of topological phases, such as topological insulators (TIs) \cite{kANE_TI1,Bernevig:2006_S,Konig:2007_S,Fu_TI,ABRMP,hasan2010colloquium,qi2011topological}, topological crystalline insulators (TCIs) \cite{fu_TCI,Bansil_TCI,Buzko_TCI,tanaka2012experimental},  Dirac semimetals (DSMs) \cite{young2012dirac,liu2014stable,armitage2018weyl,wadge2021electronic}, Weyl semimetals (WSMs) \cite{xu2015discovery,xu2015discovery1,lv2015experimental,MWSM2019}, and nodal-line semimetals (NLSMs) \cite{fang2015nodal,fang2016nodal,chang2019realization,Lei2020PRB,SaddleBS}. TIs and TCIs are bulk insulators whose nontrivial band topology, enabled by band inversions, gives rise to conducting surface states and quantized physical observables. WSMs and DSMs are three-dimensional (3D) semimetals with linear band-crossing points near the Fermi level, the so-called Weyl and Dirac points, respectively. Weyl points are inherently stable due to a quantized topological charge or chirality, which requires the breaking of inversion or time-reversal symmetry, whereas the Dirac points can only be stabilized by enforcing additional spatial symmetries in time-reversal and inversion symmetric environments. One of their hallmarks is the presence of conducting surface states that form open Fermi arcs connecting the bulk nodal points on the surface of the material. In NLSMs, the nodal points form closed loops in the bulk Brillouin zone (BZ) giving rise to characteristic drumhead states on their surfaces.  Typically, topological semimetals, such as WSMs and NLSMs, appear as intermediate phases between two topologically distinct insulating phases. The breaking of the time-reversal symmetry by introducing magnetism leads to a new variety of topological phases like the quantum anomalous Hall (QAH) phase \cite{liu2008quantum,Yu61:2010_Sci,chang2013experimental} and the axion-insulator phase \cite{Pournaghavi2021Realization,mogi2017magnetic,xu2019higher,Islam2022MnTe}.

Another appealing research direction concerns designing materials with carriers residing in flat bands which can support new correlation-driven collective phases. A comprehensive catalogue of compounds with flat bands near the Fermi energy has recently been completed \cite{regnault2021catalogue}. In the case of metals, the familiar Mott-Hubbard physics is expected. In non-metallic compounds, a key challenge is how to introduce carriers avoiding the Anderson-Mott metal-to-insulator transition, as the role of localization increases when the kinetic energy of the carriers is reduced. In 2D systems, modulation doping or gating introduces carriers without enhancing disorder, the case of flat bands in twisted 2D flakes \cite{Cao:2018_N}. Some of the present authors have recently demonstrated theoretically that 3D flat bands are expected in systems exhibiting isoenergetic nodal lines under the presence of inhomogeneous strain \cite{lau2021designing}. The latter plays the role of a gauge potential quantizing the carrier spectrum into Landau-like levels in the absence of an external magnetic field.
Rhombohedral graphite and the CaAgP-class of compounds have been pushed forward as candidate materials. However, in view of the experimental realization and study of 3D flat bands, it is paramount to find more material systems with suitable properties, especially with the presence of nodal lines.

Superlattices have been used to engineer new topological phases. One of the first general examples is a superlattice made of alternating layers of magnetically doped TIs and normal insulators to trigger the appearance of a WSM phase\cite{PhysRevLett.107.127205}. Looking for topological phases in other types of superlattices, a promising material platform constitutes zinc-blende heterostructures of HgTe, HgSe and CdTe. While CdTe is a trivial insulator, bulk HgTe and HgSe are symmetry-enforced zero-bandgap semiconductors with a band inversion, which can be tuned by changing thickness, strain, and temperature \cite{Konig:2007_S,kirtschig2016surface,Kadykov:2018_PRL,Mahler19,Mahler21}. The systems made up of these materials may give rise to several different topological phases, such as a TI phase \cite{3DQHE_2011,Kozlov:2016_PRL}, a WSM phase \cite{HgTe_weyl}, or a quantum spin Hall phase in insulating 2D HgTe/CdTe quantum wells \cite{Bernevig:2006_S,Konig:2007_S,Grabecki:2013_PRB}. In the case of magnetic Mn-doping, the $k$-dependence of the $sp$-$d$ hybridization has to be taken into account to describe accurately the exchange-induced splitting of magneto-optical spectra corresponding to electron-hole excitations in various points of the BZ \cite{autieri2020momentumresolved}. Despite the inverted band structure, coupling between Mn spins is dominated by superexchange \cite{sliwa2021superexchange} rather than by interband spin polarization \cite{Yu61:2010_Sci}. It has theoretically been predicted \cite{liu2008quantum} that the breaking of time-reversal symmetry by magnetic doping would lead to a QAH phase. However, this has not yet been observed experimentally in HgTe-based TIs.\\

In this paper, we demonstrate by {\em ab-initio} computations that (001) HgTe/CdTe and HgTe/HgSe superlattices show a rich tapestry of topological phases including phases with isoenergetic nodal lines residing at the Fermi energy of the unstrained structures. The two cases investigated here are representative of the combination of trivial/topological and topological/topological superlattices.

The paper is structured as follows. In Sec.~\ref{sec:structure}, we discuss the band structure and symmetry properties of bulk HgTe, of bulk CdTe, of bulk HgSe, and their superlattices. The computational details and the numerical values of the band structure parameters are described in Appendix A and B. Section~\ref{sec:nodal-line}, discusses the nodal-line semimetal phase obtained in unstrained HgTe/CdTe superlattices. In Sec.~\ref{sec:weyl}, we continue by presenting an ideal Weyl semimetal phase in the related HgTe/HgSe short-period superlattices, before studying the effect of hydrostatic pressure and uniaxial strain on this heterostructure in Sec.~\ref{sec:pressure-and-strain}. Notably, the latter reveals another nodal-line semimetal phase. Finally, Sec.~\ref{sec:conclusions} summarizes our results and provides a brief outlook on possible research directions.

\section{Symmetries and structural properties}
\label{sec:structure}

\subsection{Bulk}

\begin{figure}
  \begin{center}
      \includegraphics[width=0.95\linewidth]{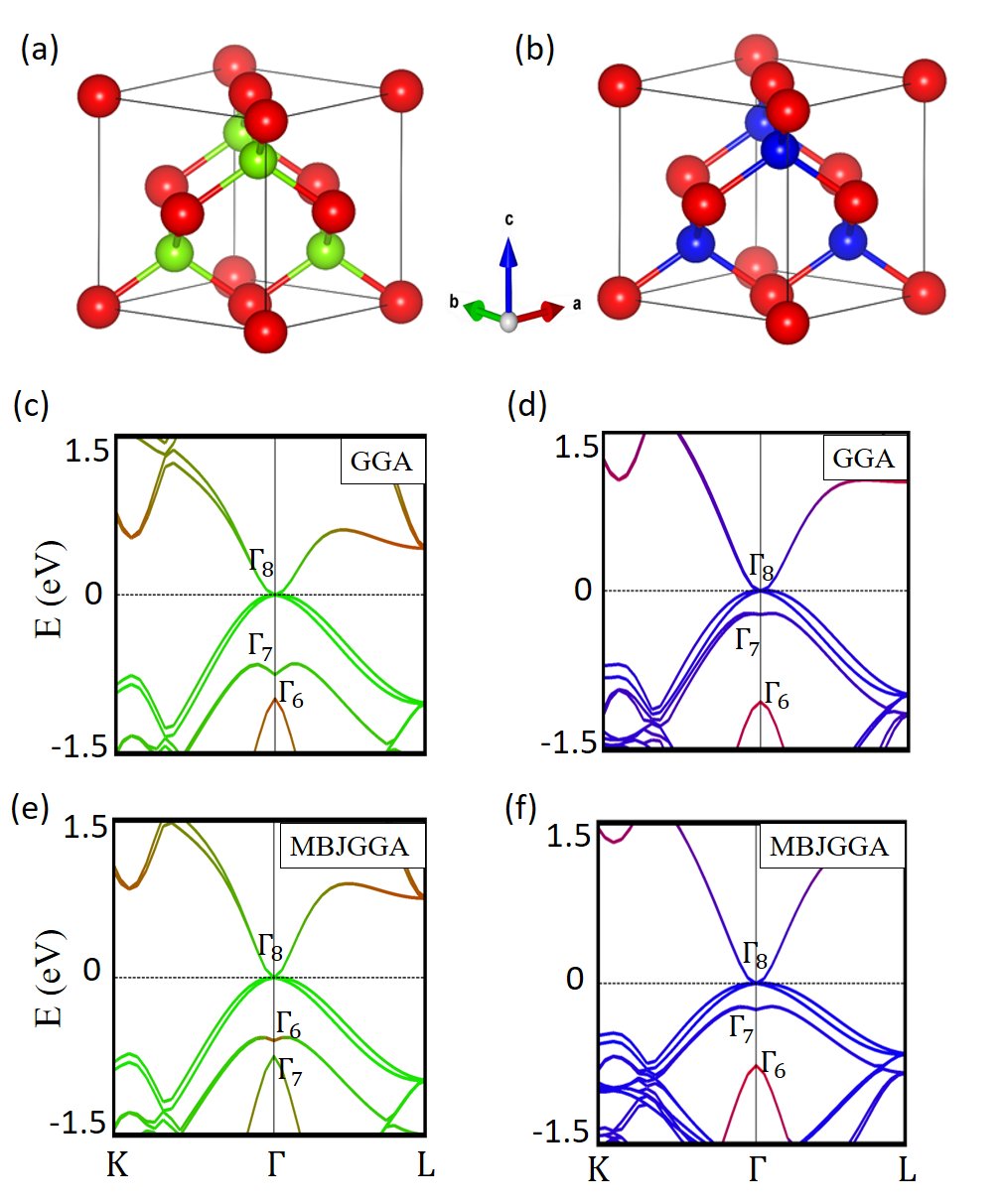}
  \end{center}
  \centering
   \caption{(a, b) Crystal structure of bulk HgTe and HgSe in the conventional unit cell. Hg, Te and Se atoms are highlighted in red, green and blue, respectively. (c-d) Band structure within GGA and (e-f) within the MBJGGA approach of bulk HgTe and HgSe, respectively. Red, green, and blue lines represent the Hg, Te and Se band characters, respectively.}
   \label{bulk}
\end{figure}

CdTe, HgTe, and HgSe belong to the space group $F\bar{4}3m$, No.\,216. They have $T_d$ symmetry with mirror planes along the (110), (101), and (011) directions. The conventional unit cell contains eight atoms: four cations and four anions as shown in ~\Cref{bulk}(a-b). The Rashba-Dresselhaus effect is present in the zinc-blende structure because of the lack of inversion symmetry\cite{dresselhaus1955spin}. 

The band structure of bulk HgTe and HgSe (Figs.~\ref{bulk}(c-f)) exhibits an inverted band ordering, dictating that HgTe and HgSe materials exhibit a nontrivial topological character~\cite{SQHE_bernevig,HgTe_band_inversion, HgTeinversion1}. The band inversion takes place between cation-$s$ and anion-$p$ bands at the Fermi level driven by two relativistic effects: on the one hand, the mass-velocity term that lowers significantly the energy of the $s$-orbitals in heavy cations and, on the other hand, spin-orbit coupling (SOC) shifting up the anion $p$ orbitals with total angular momentum $j = 3/2$\cite{Cardona2010PhysicsToday}.

The spin-orbit interaction of the anions (Se, Te) plays an important role in the band topology. The SOC splits the bands near the Fermi level into fourfold-degenerate $j={\frac{3}{2}}$ and twofold-degenerate $j={\frac{1}{2}}$ bands at the $\Gamma$ point. According to the symmetry of the wave function, the anion $p$-bands are labeled as $\Gamma_{8}$ $(j=\frac{3}{2})$ and $\Gamma_{7}$ $(j=\frac{1}{2})$, while the cation $s$-bands are labeled as $\Gamma_6$ at the $\Gamma$ point. \Cref{bulk}(c-d) shows the band structure using the generalized gradient approximation\cite{perdew1996generalized} (GGA). To reproduce the correct experimental band ordering for both HgTe and HgSe, we have further obtained band structure using the modified Becke-Johnson exchange potential together with the correlation potential scheme (MBJGGA) (see Appendix A for details). The results are shown in \Cref{bulk}(e-f).
This yields the order of energy levels at $\Gamma$ in qualitative agreement with experiments. Notably, while the band ordering between $\Gamma_6$ and $\Gamma_7$ in GGA differs from experiments, the higher-lying $p$-bands close to the Fermi level, which are relevant for the topology, are correctly obtained by both functionals. We thus resort to the GGA exchange-correlation functional to describe the topological properties of the systems in this paper.

Both HgTe and HgSe are symmetry-enforced zero-bandgap semiconductors due to the interplay between the spin-orbit splitting in the p-orbital manifold, their bulk cubic symmetry and the mass-velocity term of Hg\cite{Cardona2010PhysicsToday}. Indeed, at the $\Gamma$ point the p-orbitals split under the spin-orbit interaction into $\Gamma_8$ (four-fold degenerate) and $\Gamma_7$ (doubly degenerate) states. The remaining degeneracies are protected as long as the cubic symmetry is present. The Fermi level lies at the $\Gamma_8$ energy level producing a zero-bandgap semiconductor.
The cubic symmetry is preserved in the presence of hydrostatic pressure but it is broken in case of anisotropic strain and interface heterostructures. Once the cubic symmetry is broken, the four-fold degenerate state splits thereby removing the zero-gap semiconductor state but keeping the band inversion so that other topological phases can evolve. Additionally, the breaking of the crystal symmetry produces also the characteristic camel-back feature of the band structure.
Defining the magnitude of the band inversion as $E_g=E_p-E_s$, we have found that the band inversion strongly depends on the volume through the crystal field effect: by compressing the volume, $E_s$ increases more than $E_p$ because the $s$-orbital of the Hg atoms is isotropic and is, therefore, strongly affected by the crystal field of the four Te (Se) atoms. As a consequence, a compression of the volume reduces $E_g$ thereby pushing the system towards the trivial insulating phase.

\subsection{Superlattice}
\begin{figure}{r}
  \vspace{-20pt}
  \begin{center}
       \includegraphics[width=.95\linewidth]{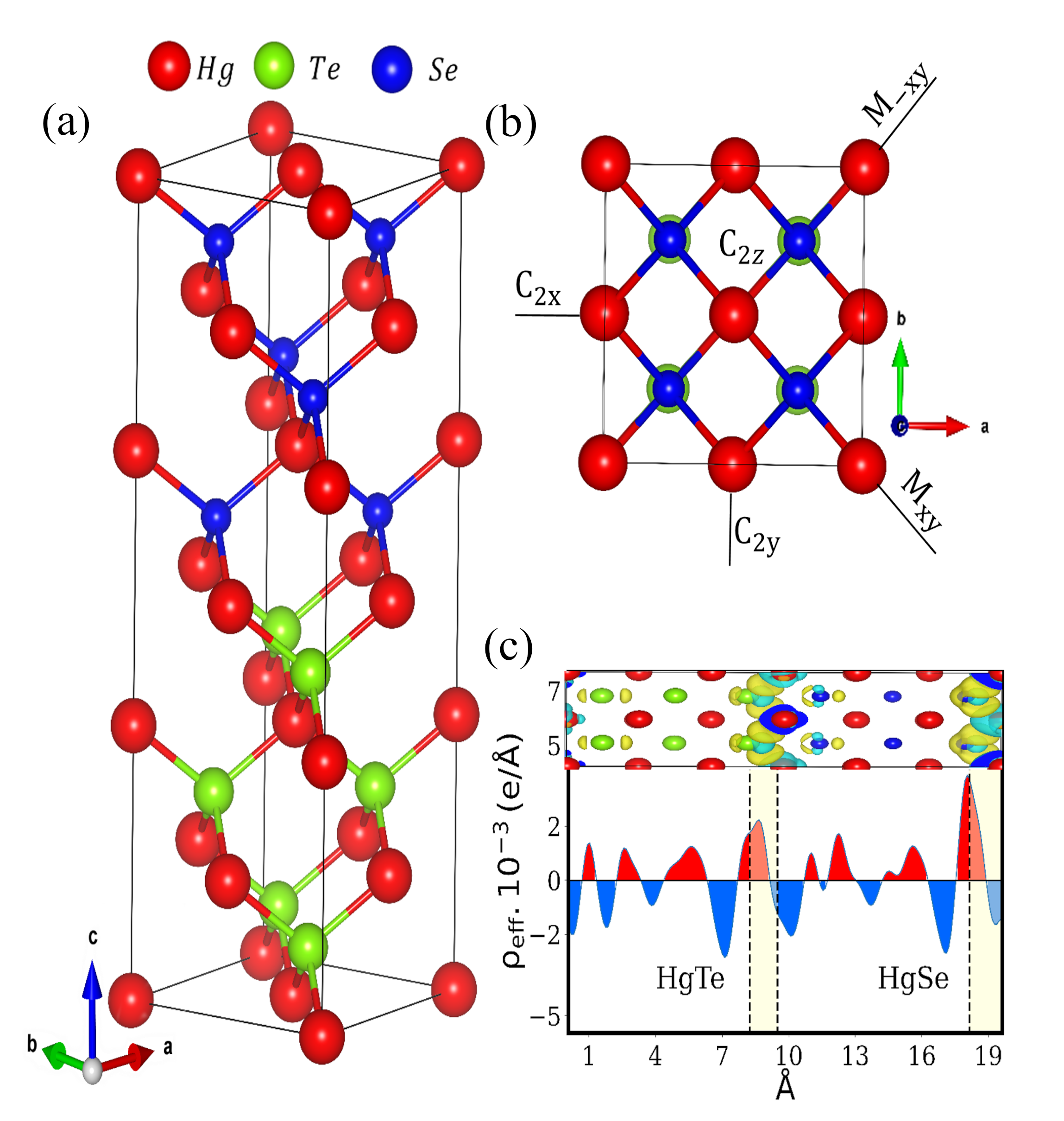}
 \end{center}
   \centering
   \caption{(a) 3D HgTe/HgSe superlattice constructed along the (001) direction. (b) Top view of the crystal structure with symmetries present in the heterostructure indicated. (c) Isosurface of the charge difference in real space (top panel) and linear charge difference $\rho(z)$ (bottom panel). We have labelled the HgTe and HgSe regions, the dashed vertical lines represent the interface layers between HgTe and HgSe.}
   \label{heterostruc}
\end{figure}

The interface between two materials couples different degrees of freedom giving rise to emergent phases such as 2D electron gases, superconductivity, proximity effects\cite{Autieri2014NJP,PhysRevB.85.075126,Roy2015}, exotic exchange bias\cite{Amitesh2014APL,Hausmann2017}, anisotropic metal-insulator transitions\cite{vanthiel2020coupling} or a sign-tunable anomalous Hall effect\cite{PhysRevLett.127.127202}. In particular, the interfaces between zincblende compounds and their superlattices have been intensively studied to generate new electronic and topological phases\cite{PhysRevLett.114.096802,PhysRevMaterials.5.084204}.\\

We study short-period HgTe/CdTe and HgTe/HgSe superlattices and reveal the emergence of various topological phases. For the sake of brevity, we only discuss the structure and symmetry properties of HgTe/HgSe, but the same statements are valid also for HgTe/CdTe. The 3D superlattice of HgTe/HgSe represents a heterostructure composed of alternating phases of two dissimilar topological semimetals. We consider the case with the same number $n$ of layers for both phases [The short-period superlattices with a small value of $n$ are considered because long-period superlattices are expected to recover the properties and phases of bulk HgTe.]. We further have to distinguish between even and odd values of $n$ in the (HgTe)$_n$/(HgSe)$_n$ heterostructures. For $n$ even, the directions (110) and ($\bar{1}$10) are equivalent. On the contrary, for an odd number of layers the (110) and ($\bar{1}$10) directions become equivalent.

We expect that the interplay between the breaking of the crystal symmetries and the full spin-orbit coupling favors a rich topological phase diagram. To break the crystal symmetries, we have constructed the superlattice by alternating three layers of HgTe and of HgSe, i.e., (HgTe)$_3$/(HgSe)$_3$, along the (001) direction. The heterostructure comprises three conventional zinc-blende unit cells along the $c$-axis and, therefore, we assume for the out-of-plane lattice constant of the supercell $c_{\text{SL}}=3a_{\text{SL}}$, where $a_{\text{SL}}$ is the in-plane lattice constant. As in the bulk, each anion (cation) is tetrahedrally coordinated by four nearest-neighbor cations (anions).
When cutting through the structure along the $z$-direction, one finds subsequent layers of one atomic thickness consisting of only anions and only cations, respectively [see ~\Cref{heterostruc}(a)].
The six cationic layers are composed of Hg atoms, while there are three anionic layers containing Te and three anionic layers containing Se in each superlattice period.
Both compounds have the same zinc-blende space group, but in the heterostructure with an odd number of layers the directions (110) and ($\bar{1}$10) are now inequivalent.
This results in a lower symmetry which could favor the possibility of Weyl and nodal-line phases\cite{ruan2016ideal}.
Moreover, the interface between zinc-blende materials can be used to manipulate the band gap and, therefore, the topological properties of the whole heterostructure.
We will show that these 3D superlattices have electronic properties similar to the HgTe class of materials under strain \cite{HgTe_weyl}, since they preserve the C$_2$ symmetry. However, the different periodicity along the $z$-axis lends the superlattices a tetragonal symmetry.

Our heterostructure belongs to the space group $P\bar{4}m2$ (D$_{2d}$), No.\,115, which is a body-centered tetragonal crystal structure with lattice parameters $a_{\text{SL}}$ and $c_{\text{SL}}$. By design, the point-group symmetry of the heterostructure is reduced from $T_d$ to $D_{2d}$ with respect to the bulk compounds.
Bulk HgTe and bulk HgSe have lattice constants $a_{\text{HgTe}}=6.46$\,{\AA} and $a_{\text{HgSe}}=6.09$ {\AA}, respectively. The in-plane lattice constant of the supercell depends on the lattice constants of the constituent compounds following Vegard's law.
Without strain, we assume the in-plane lattice constant of the supercell to be $a_{\text{SL}}=6.27$\,{\AA}, which is the rounded average between the experimental HgTe and HgSe lattice constants and therefore expected to be close to the experimental value for this superlattice.

The heterostructure has three two-fold rotational symmetries along the $x$, $y$ and $z$ directions (C$_{2x}$, C$_{2y}$, and C$_{2z}$), and two mirror symmetries (M$_{xy}$ and M$_{x\bar{y}}$), as shown in ~\Cref{heterostruc}(b). Similar to bulk HgTe, the (HgTe)$_3$/(HgSe)$_3$ heterostructure lacks inversion symmetry. Nevertheless, it preserves time-reversal symmetry. Notably, the symmetry properties of the heterostructure are similar to strained HgTe \cite{HgTe_weyl} and some chalcopyrites \cite{S4weyl1,S4weylhigh}.
The presence of time-reversal symmetry together with the mirror symmetries and the $C_{2z}$ symmetry give rise to doubly degenerate bands along the $\Gamma$-Z direction, so-called Kramers nodal lines~\cite{Xie2021}, while the band structure has nondegenerate bands along all other directions as discussed below.
This implies a constraint on the topology of the Fermi surfaces similar to the effect of nonsymmorphic symmetries \cite{Cuono2019}.

Interfaces between two materials generally give rise to a charge transfer between the layers. To assess this effect,  we define the linear charge density $\rho(z)$ as the number of electrons summed over the $xy$ plane per unit length such that $\int\rho(z)dz$ is equal to the number of electrons. We have performed density functional theory calculations for the entire heterostructure and for separate slabs containing only HgTe and only HgSe with the same geometry as the heterostructure. Subsequently, we have considered the difference between the respective linear charge densities:
\begin{equation}
\Delta\rho(z)=\rho_{\text{HgTe/HgSe}}(z)-\rho_{\text{HgTe}}(z)-\rho_{\text{HgSe}}(z).
\end{equation}
The integral of this quantity is zero by construction. We find that a negligible amount of charge on the order of $10^{-3}$ electrons is accumulated near the interfaces as shown in ~\Cref{heterostruc}(c). These results infer that there is no significant charge transfer between HgTe and HgSe in the 3D superlattice.
This confirms our expectations as the two compounds have similar electronegativities.
In the HgTe/HgSe heterostructure there is one kind of interface, namely Te/Hg/Se. However, the heterostructure can be cut in two regions in different ways. We cut the heterostructure in two different ways at Te/Hg and Hg/Se. For this reason, we note that there is a slight asymmetry between the two interfaces as seen in~\Cref{heterostruc}(c). 
These results clearly demonstrate that there is no considerable charge transfer at the interface. 

In the following sections, we show the appearance of a nodal-line semimetal phase in short-period superlattices of HgTe/CdTe and of an ideal Weyl semimetal phase in HgTe/HgSe superlattices.
Moreover, we will investigate the latter under hydrostatic and uniaxial pressure.

\section{Nodal-line semimetal in H\lowercase{g}T\lowercase{e}/C\lowercase{d}T\lowercase{e} superlattices}
\label{sec:nodal-line}

\begin{figure}[t]
  \centering
  \includegraphics[width=0.95\linewidth]{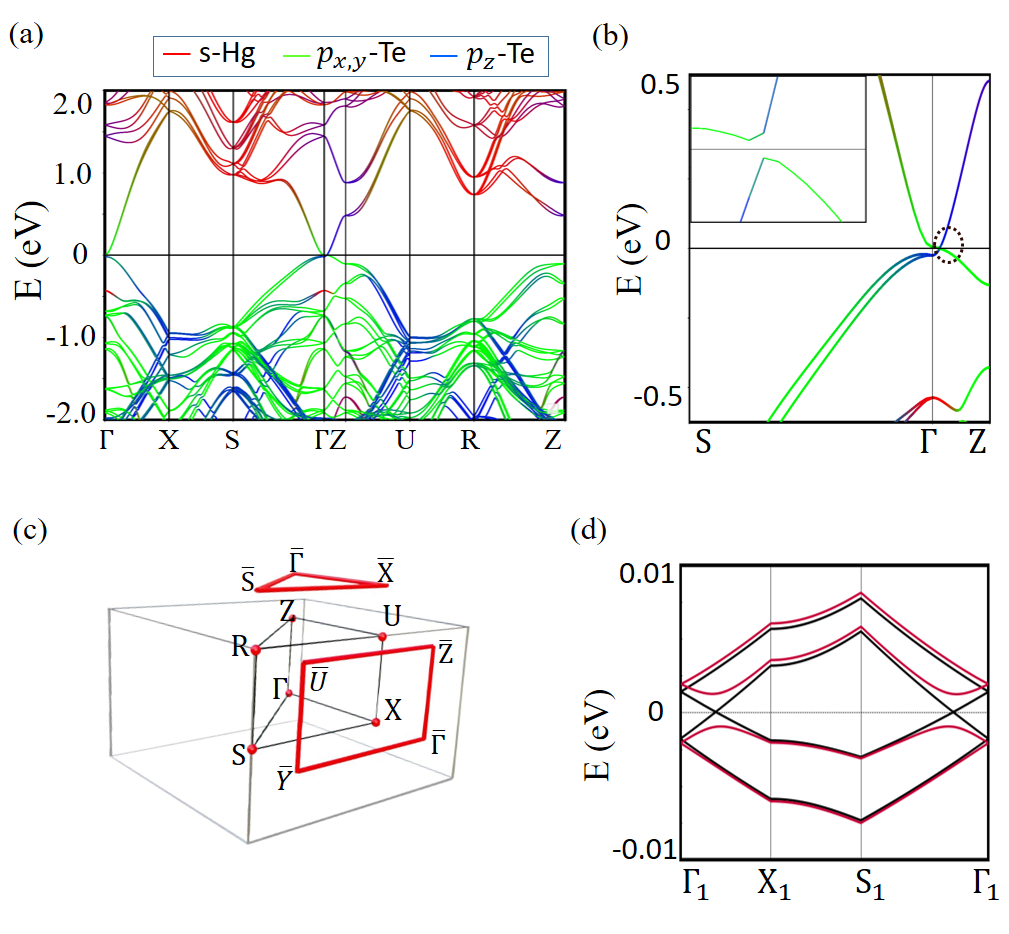}
  \caption{(a) Orbital-resolved band structure of (HgTe)$_3$/(CdTe)$_3$ superlattice in the full Brillouin zone (BZ) and (b) enlarged view along the S-$\Gamma$-Z high-symmetry path.
  The inset shows that the system is gapped along the $\Gamma$-Z direction. (c) Schematic diagram of the full BZ (black solid lines) with (100) and (001) BZ surface projections (red solid lines). (d) Band structure with nodal points at $k_z$=0.020 \AA$^{-1}$(black) and gapped at $k_z$=0.021 \AA$^{-1}$(red). These four bands are a set of bands isolated from the rest of the band structure.}
  \label{fig : CdTe_L6.60}
\end{figure}

\begin{figure}[t]
  \centering
  \includegraphics[width=0.95\linewidth]{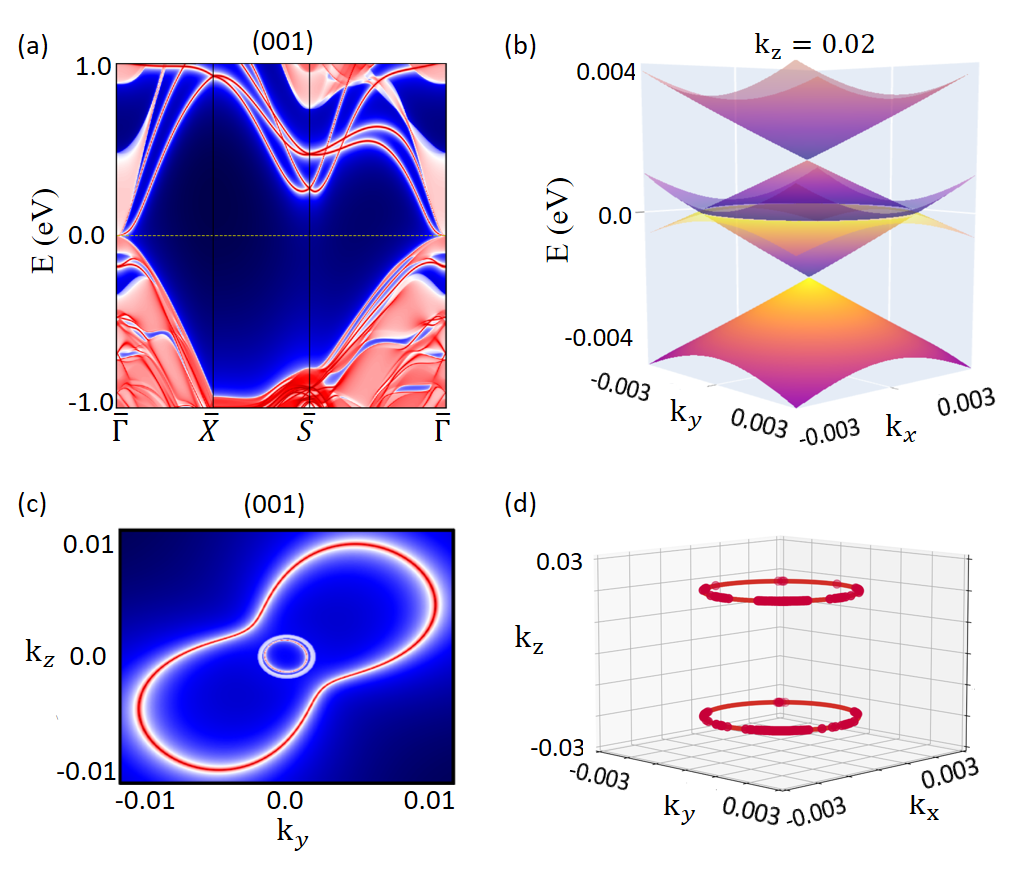}
  \caption{(a) Surface band structure and (c) associated Fermi band contours of (001) Te-terminated surface of (HgTe)$_3$/(CdTe)$_3$ superlattice. (b) Band structure of the infinite superlattice in the $E-k_x-k_y$ space at a fixed $k_z$=0.02 \AA$^{-1}$. (d) Nodal lines of the infinite superlattice resolved at the $k_z$=$\pm$0.02 \AA$^{-1}$ planes.}
  \label{fig : CdTe_L6.60_surface}
\end{figure}

We now discuss the band structure and the Fermi surface of the unstrained (HgTe)$_3$/(CdTe)$_3$ superlattice. This superlattice is feasible experimentally due to the presence of only one type of chalcogen atoms as assessed in the earlier quantum well experiments. \Cref{fig : CdTe_L6.60}(a) shows the band structure along high-symmetry lines of the full BZ. At the $\Gamma$ point, the s-band of Hg is at $-0.5$ eV with respect to the Fermi level, which is similar to bulk HgTe. The bands close to the Fermi level are dominated by the $p$-orbitals of the Te atoms. An enlarged view of the band structure along the S-$\Gamma$-Z high-symmetry path in \Cref{fig : CdTe_L6.60}(b) reveals a band inversion between the $\ket{j,j_z}=\ket{3/2,\pm1/2}$ bands and the $\ket{j,j_z}=\ket{3/2,\pm3/2}$ bands of the Te atoms.
Looking at the orbital weight of these bands, we find that $\ket{3/2,\pm3/2}$ are made up only of p$_x$ and p$_y$ orbitals, while $\ket{3/2,\pm1/2}$ contain one third of in-plane orbitals and two thirds of p$_z$. We make similar observations for the other type of superlattice considered in this article (see below).
While the band structure is gapped along the $\Gamma$-Z direction, we find that the system features two isoenergetic, circular nodal-lines with the same radius in planes parallel to the $k_x$-$k_y$ plane at $k_z^*=\pm 0.02$~\AA$^{-1}$. ~\Cref{fig : CdTe_L6.60}(d) shows the corresponding bands along two paths of the form $\Gamma_1=(0,0,k_z) \rightarrow X_1=(0.005\frac{\pi}{a}, 0,k_z) \rightarrow S_1=(0.005\frac{\pi}{a}, 0.005\frac{\pi}{b},k_z) \rightarrow \Gamma_1$, one with $k_z=k_z^*$ crossing one of the nodal lines and one with $k_z= 1.05\,k_z^*$. Remarkably, the nodal lines are isoenergetic with lack of energy dispersion and lie at the Fermi level.

~\Cref{fig : CdTe_L6.60_surface}(a) shows the surface band structure with Te-terminated (001) surface. The associated Fermi contours has two circular features close to $\bar{\Gamma}$, as shown in ~\Cref{fig : CdTe_L6.60_surface}(c), one of which corresponds to the projection of the bulk nodal-lines into the surface BZ. The other circular feature together with a larger bow-tie shaped Fermi line are made up of surface states. In ~\Cref{fig : CdTe_L6.60_surface}(b), we present the superlattice band structure at fixed $k_z=k_z^*=0.02$ \AA$^{-1}$, where the circular nodal line is visible at the Fermi level. The Fermi surface of the superlattice is shown in ~\Cref{fig : CdTe_L6.60_surface}(d) consisting of two circular lines located on the $k_z = \pm k_z^* = \pm$0.02 \AA$^{-1}$ planes. The superlattice of HgTe/CdTe thus realizes a nodal-line semimetal phase.\\

The presence of these nodal lines can be understood from looking at bulk HgTe.
Away from the Fermi level, bulk HgTe features a network of bow-tie shaped nodal lines protected by six symmetry-related mirror planes \cite{SuppMatRuan2016}.
The application of uniaxial strain along the c-axis breaks all mirror symmetries except $M_{xy}$ and $M_{x\bar{y}}$.
Consequently, all nodal lines except the ones in the two remaining mirror planes are gapped out.
However, the latter gradually shrink to points and gap out already for small strain.
Finally, the only remaining features are eight $C_2T$ symmetry-protected point nodes in the $x$-$z$ and $y$-$z$ planes, which are located along their intersections with the broken mirror planes.
In particular, they lie on some of the previous but no longer protected nodal lines.
In our heterostructure, the role of the external strain is played by the interface between the two dissimilar materials leading to a similar picture.
Remarkably, the band structure parameters of the short-range superlattice with trivial CdTe lead to a situation where the nodal lines containing the $C_2T$ protected point nodes survive and move to the Fermi level. 
Notably, while these nodal lines are not symmetry protected, the gap opened at the crossings points are negligible and beyond our computational resolution. This situation is comparable to many 3D Dirac semimetal materials. These states lie close to the Fermi level, forming almost dispersionless flat bands as seen in Fig.~\ref{fig : CdTe_L6.60_surface}(d).\\


We note that we find a pair of nodal lines also in the (HgTe)$_4$/(CdTe)$_4$ superlattice (see Appendix C) despite the different symmetries. However, the nodal lines tend to shrink as we increase the   of the superlattice. We also note that in 2D quantum wells there is a critical thickness for the topological phase of HgTe. However, in the case of the superlattice we expect this critical thickness to be shorter due to the repetition of HgTe. Finally, we find that the considered HgTe/CdTe superlattice becomes a trivial insulator when applying compressive hydrostatic pressure.

\begin{figure}[b!]
  \centering
  \includegraphics[width=0.95\linewidth]{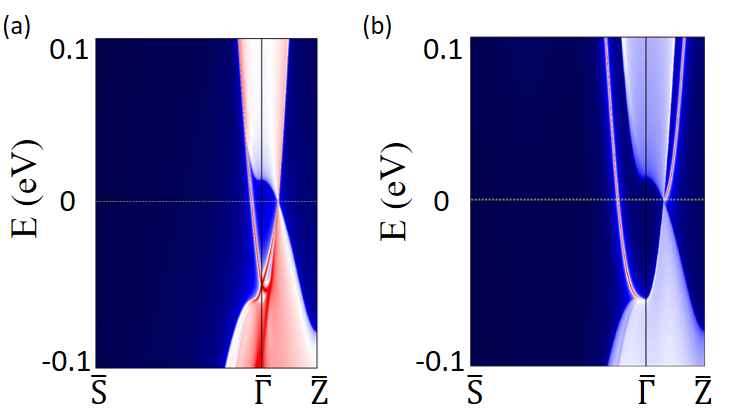}
  \caption{Band structure of the slab with (100) orientation for the (a) Hg-terminated surface and (b) Te-terminated surface.} 
  \label{fig : 100_differenttermination}
\end{figure}

\begin{figure*}[!ht]
  \centering
  \includegraphics[width=.99\linewidth]{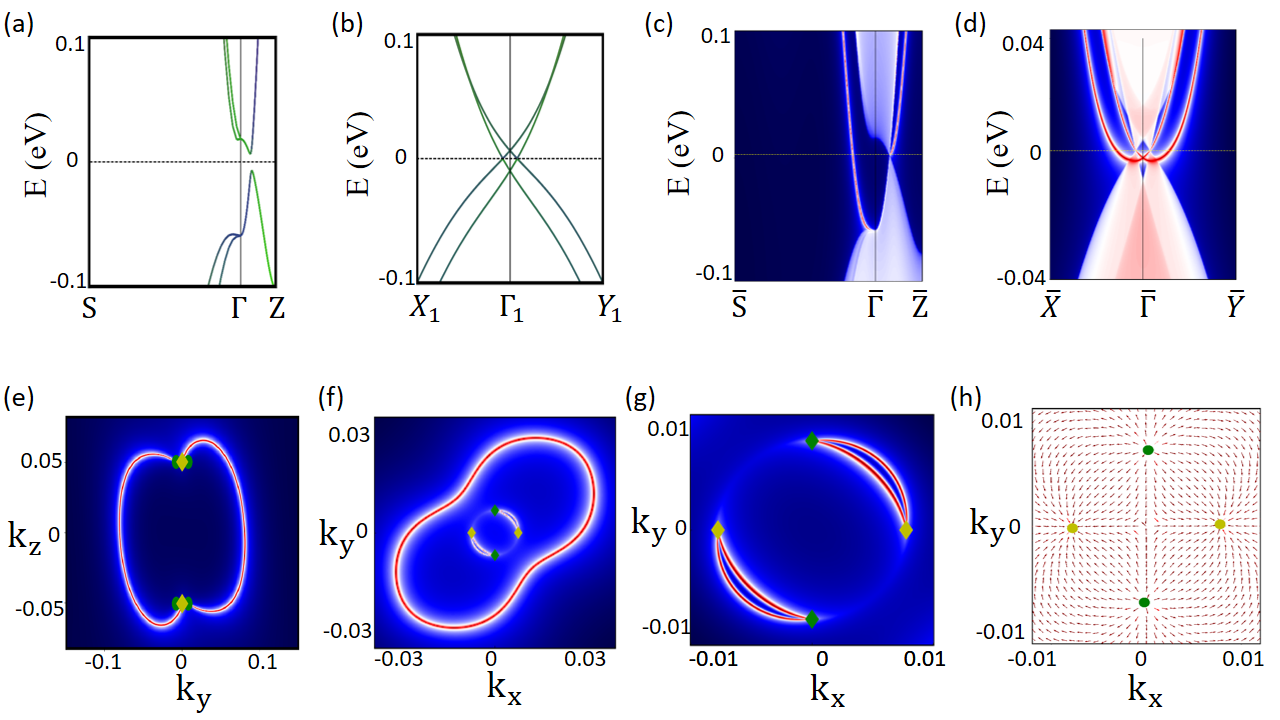}
  \caption{Band structure of the unstrained HgTe/HgSe infinite superlattice (a) along the high-symmetry lines S-$\Gamma$-Z. (b) Band structure in the plane of the Weyl points at $k_z=k_z^*=0.0506$\,{\AA$^{-1}$}. Band structure of slabs with (c)  (100) surfaces and (d)  (001) surfaces. Fermi surfaces of the slabs with (e) (100) surfaces and (f) (001) surfaces. (g) Magnification of panel (f) close to $\bar{\Gamma}$. (h) In-plane component of the Berry flux at the $k_z=k_z^*$ plane containing four Weyl points. The Weyl points at $(\pm {{k_\parallel}^*},0,\pm {k_z^*})$ have chirality -1  (yellow circle marker) while the Weyl points at  $(0,\pm {k_\parallel}^*,\pm {k_z^*})$ have chirality +1 (green circle marker). Diamond markers indicate chirality -2 (yellow) and +2 (green).}
  \label{fig : L6.27}
\end{figure*}

\section{Ideal Weyl semimetal in H\lowercase{g}T\lowercase{e}/H\lowercase{g}S\lowercase{e} superlattices}
\label{sec:weyl}

We now consider the unstrained HgTe/HgSe superlattice. We find that the details of the surface states depend on the type of surface termination. \Cref{fig : 100_differenttermination} (a) and (b) show the surface band structure of the Hg- and Te-terminated surfaces, respectively. We observe surface states connecting the valence and the conduction bands indicating the topological nature of the system. For the Hg-termination, the surface Dirac point is at 0.06 eV below the Fermi level. On the contrary, the Dirac point is buried in the bulk bands for the Te-termination of the slab with (100) orientation. For the sake of brevity,  we consider chalcogen atomic terminations to present our following results.  

\Cref{fig : L6.27}(a) displays the band structure of the infinite superlattice along the high-symmetry lines S-$\Gamma$-Z close to the Fermi level. It constitutes an inverted band gap of 13.1\,meV on the $\Gamma$-Z line. A careful exploration of states near the band anticrossing points reveals linear band crossing typical of Weyl semimetals along the path X$_1$-$\Gamma_{1}$-Y$_1$, as shown in ~\Cref{fig : L6.27}(b). We define $\Gamma_{1}$=(0,0,$k_z^*$), X$_1=(0.1\pi/a,0,k_z^*)$ and Y$_1=(0,0.1\pi/a,k_z^*)$, with $k_z^*=0.0506$\,{\AA$^{-1}$}.
We find a total of eight symmetry related Weyl points constrained by the presence of $C_{2x}T$ and $C_{2y}T$ symmetries with respect to the planes at $k_x$=0 and $k_y$=0\cite{ruan2016ideal}. Their positions in the BZ are $(0,\pm k_\parallel^*, \pm k_z^*)$ for the four Weyl points with chirality $+1$ and $(\pm k_\parallel^*, 0, \pm k_z^*)$ for the other four with chirality $-1$, where we have defined $k_\parallel^*=0.0077$\,{\AA$^{-1}$}. Interestingly, these Weyl nodes appear at the Fermi level, showing that the superlattice structure realizes ideal Weyl semimetal phase in their pristine state in contrast to HgTe~\cite{HgTe_weyl}. 

The orbital-resolved band structure further reveals that the $p$-bands of Se and Te dominate near the Fermi level. In addition to the band inversion present in bulk HgTe and HgSe, also in this superlattice we find a new band inversion between $\ket{3/2,\pm3/2}$-Se and $\ket{3/2,\pm1/2}$-Te orbitals at the $\Gamma$ point which is essential for the creation of the Weyl phase.
The system thus has multiple band inversions at the $\Gamma$ point. The symmetry is reduced to D$_{2d}$ due to the interface. In general, the symmetry reduction creates a camel-back-shaped band in the valence bands and an inverse camel-back structure in the conduction bands along the $\Gamma$-Z high-symmetry line. The same camel-back feature has been found for bulk HgTe in the presence of doping or strain, which also breaks the bulk symmetry.

A material with inversion and time-reversal symmetry would have doubly degenerate bands due to Kramers's degeneracies. The splitting between the spin-orbit split bands gives a measure of the bulk inversion asymmetry (BIA)\cite{ruan2016ideal}. To create Weyl points in HgTe materials class, both BIA and symmetry reduction to $D_{2d}$ are needed. The parameter $\alpha$ governing the first-order term in \textbf{k} of the BIA has been estimated to be small in HgTe ($\alpha=0.208$ {\AA} eV) \cite{ruan2016ideal}. Due to the small value of ${k_\parallel}^*$, the third-order term in \textbf{k} can be neglected. For the considered superlattice, the spin-orbit split bands reported in ~\Cref{fig : L6.27}(b) cross each other in the conduction band.
Curiously, at this crossing point, the BIA is accidentally zero. Later, this observation will serve useful in understanding the progression of topological phases under applied external strain.

One of the hallmarks of WSMs is the presence of topologically protected Fermi arcs on the surface of the material. To confirm the topological nature of the 3D superlattices, we have calculated the surface electronic states and Fermi arcs for the (100) and (001) surface orientations, using the surface notation of the conventional unit cell as shown in \Cref{heterostruc}(a). The (010) surface is equivalent to the (100) surface.

\begin{figure}{!b}
  \begin{center}
       \includegraphics[width=.95\linewidth]{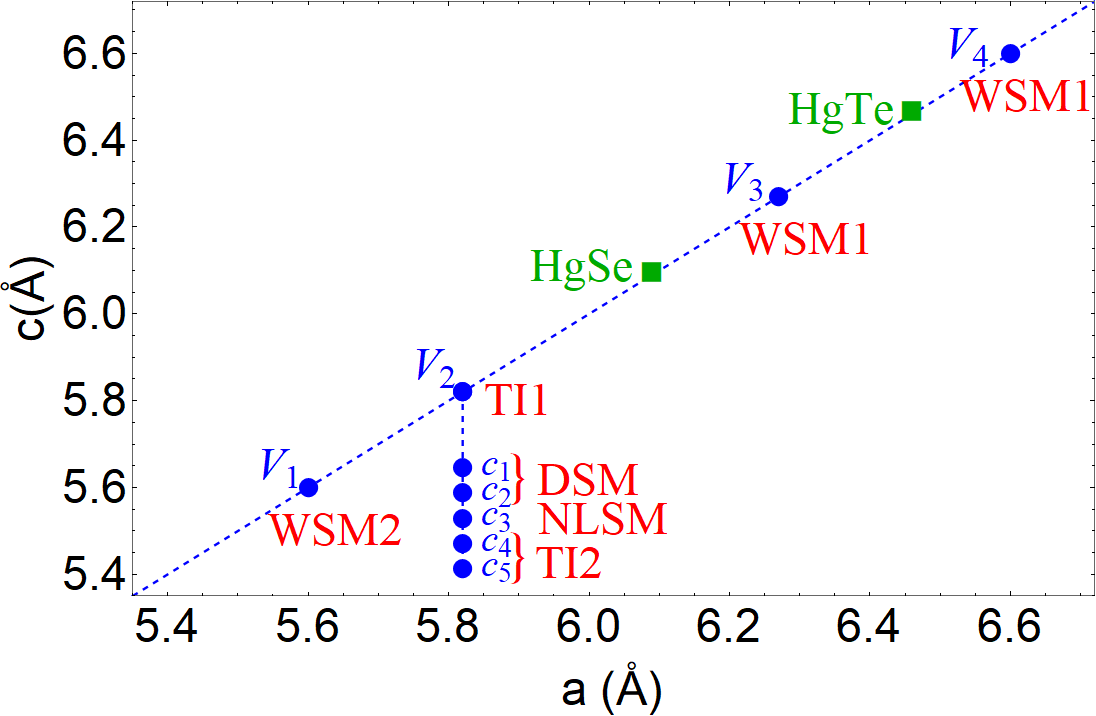}
  \end{center}
   \caption{Overview of topological phases in the HgTe/HgSe superlattices as a function of the lattice constants $a$ and $c$: blue dots show the considered configurations of lattice constants. We denote different volumes resulting from hydrostatic pressure as $V_i$, where $V_3$ corresponds to the unstrained superlattice. In the case of uniaxial strain, we denote the compressed c-axis values with fixed in-plane lattice constant as $c_i$. The experimental values of bulk HgTe and HgSe are indicated by green squares. We have further indicated the associated topological phases for each point.}
   \label{diagram}
\end{figure}

The band structure of the slab for the (100) surface orientation along high-symmetry lines is shown in \Cref{fig : L6.27}(c).
We find surface states connecting the valence and the conduction bands confirming the topological nature of the system.
Moreover, we observe a gapless point at the coordinates $(0,k_z^*)$ corresponding to Weyl points projected into the 2D BZ of the slab.
The Fermi surface for the slab with surface orientation (100) is shown in \Cref{fig : L6.27}(e).
It has six gapless points, four points with projected coordinates $(\pm {k_\parallel}^*,\pm k_z^*)$ and monopole charge +1 and two points with coordinates $(0,\pm k_z^*)$ and monopole charge -2.
The latter is the result of two Weyl points of charge -1 being projected onto the same point in the surface BZ.
We observe two large Fermi arcs connecting each one of the -2 monopole charges with
one of the -1 monopole charges.
The other two short Fermi arcs, which are expected to be present between the monopole charges with -2 and the remaining two monopole charges with -1, are not clearly resolved due to the small distances between the Weyl points for this surface orientation.

The band structure for the slab with (001) surface orientation along high-symmetry lines is shown in \Cref{fig : L6.27}(d) featuring two gapless points.
We also present the corresponding Fermi surface and its magnification around the Weyl-point projections in \Cref{fig : L6.27}(f) and \Cref{fig : L6.27}(g), respectively.
The Weyl points with monopole charge +1 are pairwise projected to the points $(0,\pm{{k_\parallel}^*})$, while the Weyl points with monopole charge of -1 are pairwise projected to $(\pm{{k_\parallel}^*},0)$.
The resulting four nodal points have an effective monopole charge of $\pm 2$ giving rise to two Fermi lines emanating from each of them.
In this way, they form four open Fermi arcs between them giving rise to two separate intra-connected pairs of Weyl nodes.
We also observe one large closed Fermi line encompassing the surface projections of the Weyl points around $\bar{\Gamma}$.
Notably, this Fermi line is disconnected from the Weyl points and originates from the vicinity to a topological insulator phase that would form after pairwise annihilation of the Weyl points.
The remnant surface Dirac cone can be viewed as coming from the bulk band inversion, while the band-inversion generated by the $p$-orbitals produces the Weyl points.
The resulting coexistence of Dirac cone and Weyl points has been described in the literature~\cite{LKBO2017}. Furthermore, we have calculated the Berry flux in the $k_z=k_z^*$ planes containing the Weyl points. The projections of the Berry flux into the $x$-$y$ plane are visualized in ~\Cref{fig : L6.27}(h). We see how the Weyl points with chirality $+1$ act as sources of Berry flux while the Weyl points with chirality $-1$ act as sinks. The Weyl points of opposite chirality are well separated by an in-plane distance of 1.5\% of the reciprocal lattice constant. Since there are no trivial states at the Weyl nodes' energy, the Weyl phase in this 3D multilayer superlattice could be unambiguously detected experimentally.


\begin{figure*}[!ht]
  \includegraphics[width=0.95\linewidth]{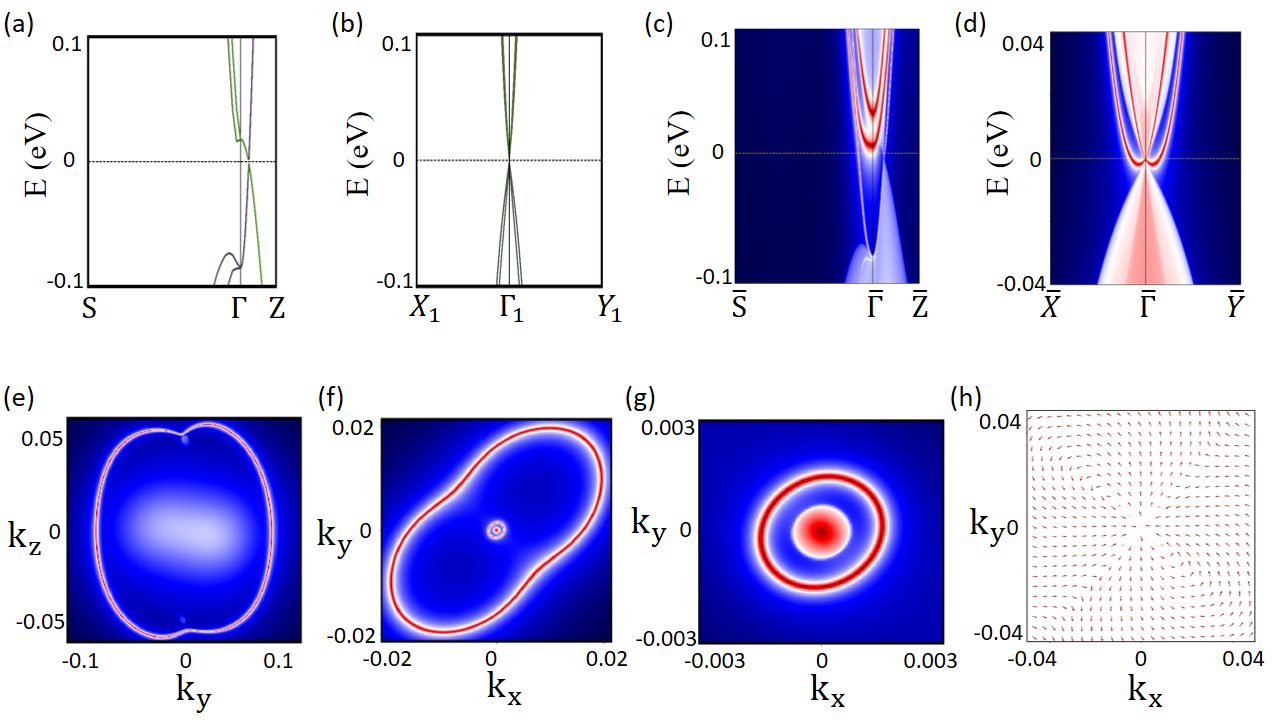}
  \caption{Band structure of the HgTe/HgSe superlattice at the volume $V_2$ (a) along the high-symmetry lines S-$\Gamma$-Z. (b) Band structure in the $k_x$-$k_y$ plane at $k_z=k_z^*=0.0360 $\AA$^{-1}$. Band structure of slabs with (c) (100)  and (d) (001)  surfaces.  Fermi surfaces of the slabs with (e) (100) surfaces and (f) (001) surfaces. (g) Magnification of panel (f). (h) In-plane component of the Berry flux at the $k_z=0$ plane.}
  \label{fig : L5.82}
\end{figure*}

\section{Hydrostatic pressure and uniaxial strain in H\lowercase{g}T\lowercase{e}/H\lowercase{g}S\lowercase{e} superlattices}
\label{sec:pressure-and-strain}

\subsection{Topological properties as a function of hydrostatic pressure}

We present the computational results on the evolution of topological phases in the HgTe/HgSe superlattices with applied hydrostatic pressure. In particular, we have performed calculations for different volumes, labelled $V_i$, $i=1,2,3,4$ as shown in \Cref{diagram}, where volume $V_3$ with $a_{\text{SL}}$=6.27\,{\AA} corresponds to the unstrained superlattice discussed in the previous section. For volumes $V_1$ and $V_2$, we have used the lattice constants $a_{\text{SL}}$ = 5.60\,{\AA} and 5.82\,{\AA}, respectively, whereas volume $V_4$ has a lattice constant of $a_\text{SL}=6.60$\,{\AA}.

For volume $V_4$, we obtain a Weyl phase similar to the one reported for volume $V_3$ above (see Appendix C). We denote the Weyl phase found for volumes $V_3$ and $V_4$ as WSM1. Comparing $V_3$ and $V_4$, we find that with increasing pressure  $k_{\parallel}^*$  is reduced while $k_z^*$ remains almost constant. The Weyl points move closer to the points (0,0,$\pm k_z^*$) where they eventually annihilate each other as we show in the next subsection. We highlight that the considered heterostructures allow to obtain the Weyl phase even in absence of strain, which is an advantage over the bulk phases of HgTe and HgSe.

For the volume $V_1$, we observe a different WSM phase which we denote WSM2. The WSM phases WSM1 and WSM2 are separated by a small-gap topological insulator, which we denote TI1. In the following, we discuss the phases TI1 and WSM2 in more detail.

\subsubsection{Small-gap topological insulator phase at volume $V_2$}

\begin{figure*}[!ht]
  \centering
  \includegraphics[width=0.95\linewidth]{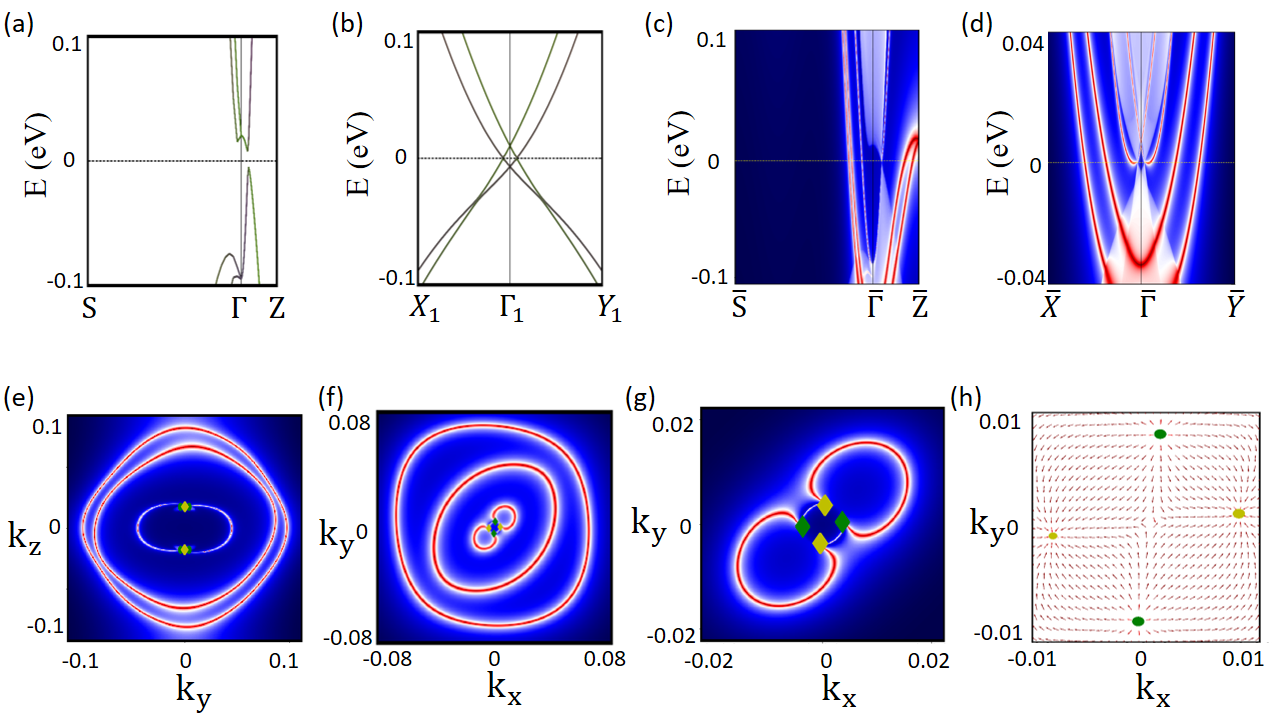}
  \caption{Band structure of the HgTe/HgSe superlattice at the volume $V_1$ (a) along the high-symmetry lines S-$\Gamma$-Z. (b) Band structure in the plane containing the Weyl points at $k_z=k_z^*=0.0353$~{\AA$^{-1}$}. Band structure of slabs with (c) (100)  surfaces and with (d) (001)  surfaces. Fermi surfaces of the slabs with (e) (100) surfaces and with (f) (001)  surfaces.  (g) Magnification of panel (f) to visualize the Weyl points and their connectivity. (h) In-plane component of the Berry flux at the $k_z={k_\parallel}^*$ plane including four Weyl points. The Weyl points at $(\pm {{k_\parallel}^*},0,\pm {k_z^*})$ have chirality -1 (yellow diamond marker), while the Weyl points at  $(0,\pm {{k_\parallel}^*},\pm {k_z^*})$ have chirality +1 (green circle marker). Diamond markers indicate chirality -2 (yellow) and +2 (green).}
  \label{fig : L5.60}
\end{figure*}

\Cref{fig : L5.82} shows the electronic properties for the compressed superlattice with volume $V_2$. Along the high-symmetry lines S-$\Gamma$-Z, the infinite superlattice band structure
has a minimum band gap of 3~meV at $(0, 0, 0.0360\, \textrm{\AA}^{-1}$) as shown in ~\Cref{fig : L5.82}(a).
The existence of a finite energy gap is further confirmed in the a very denser grid using our Wannier model. We have performed calculations along the path X$_1$-$\Gamma_{1}$-Y$_1$ with $k_z$ components shifted to $k_z=k_z^*=0.0360$ {\AA$^{-1}$}, which is presented in~\Cref{fig : L5.82}(b). Moreover, the points where the BIA splitting between the bands accidentally vanishes have now moved to the Fermi level and are now closer to the $\Gamma$ point coinciding with the position of the minima of the bulk energy gap.
This means that the BIA is tunable by the hydrostatic pressure and vanishes approximately at the Fermi level for the volume $V_2$.
Coming from the Weyl phase, the Weyl points merge and gap out at the two points $(0,0,\pm{k_z^*})$ leading to a small-gap topological insulator that we denote TI1. This small-gap TI phase can also be viewed as an approximate Dirac semimetal with Dirac points at $(0,0,\pm{k_z^*})$ since it is in proximity of a DSM phase.
This DSM phase is not topologically protected. Due to the presence of multiple band inversions, the gapped phase emerging here is a topological insulator phase.
The recombination of Weyl points at $k= (0,0,\pm{k_z^*})$ due to the accidental vanishing of the BIA is unique to the superlattices considered here. In strained HgTe, on the contrary, the merging of Weyl points happens at the $\Gamma$ point.

We have further studied the electronic surface states of slabs of the compressed superlattice to confirm the nontrivial topological nature of the system. The band structures for the surface orientations (100)  and (001) are presented in~\Cref{fig : L5.82}(c) and (d), respectively. These look similar to the corresponding band structures for the unstrained superlattice (compare to \Cref{fig : L6.27}), except that the pair of linear band crossings in the bulk continuum has now merged to a single point in~\Cref{fig : L5.82}(d).
This is in agreement with the recombination of Weyl points to a single Dirac point.
Furthermore, we still observe topological surface states connecting the conduction and valence bands indicative of the topological nature of the compressed superlattice.
A deeper insight into the structure of the surface states is gained from the corresponding Fermi surfaces for the (100) and (001) surface orientations of the slabs, as presented in \Cref{fig : L5.82}(e)-(g)
Again comparing to \Cref{fig : L6.27} for the (100) termination in~\Cref{fig : L5.82}(e) we see that the previously open Fermi arcs now form a closed Fermi line of surface states pinned to the surface projections of the infinite superlattice Dirac points at $(0,\pm k_z^*)$.
This Fermi line belongs to a single surface Dirac cone characteristic of a topological insulator.
For the (001) termination in~\Cref{fig : L5.82}(f) and~(g), we find a large closed Fermi line and two small Fermi circles centered at the $\bar{\Gamma }$.
The inner Fermi circles originate from the fusion of the Weyl points.



\subsubsection{Weyl semimetal at volume $V_1$}

Reducing the volume further to the value $V_1$, we find another WSM phase that we denote as WSM1. The corresponding Weyl points re-emerge from the critical Dirac points found for the volume $V_2$. This is due to the BIA increasing again, which accidentally vanished close to $V_2$. From the point of view of a model Hamiltonian, this can be understood considering that changes in the volume modify the ratio between the hopping parameters and the on-site energies, and therefore the BIA. \Cref{fig : L5.60}(a) shows the infinite superlattice band structure at the volume $V_1$ along the high-symmetry lines S-$\Gamma$-Z. Along these high-symmetry lines the minimum bandgap is
12.3 meV. We have also calculated the band structure along the shifted path X$_1$-$\Gamma_{1}$-Y$_1$ with $k_z$ components $k_z=k_z^*=0.0353$ \AA$^{-1}$, as shown in ~\Cref{fig : L5.60}(b).
Along this path, we find linear dispersions with gap closing points. There are a total of eight Weyl points.
Their chiralities and relative arrangement in the BZ are similar to WSM2: The positions are $(0,\pm k_\parallel^*, \pm k_z^*)$ for the four Weyl points with chirality $+1$ and $(\pm k_\parallel^*, 0, \pm k_z^*)$ for the other four with chirality $-1$, where we have $k_z^*=0.0353$\,\AA$^{-1}$ and $k_\parallel^*=0.0038$\,\AA$^{-1}$.
Differently from the Weyl semimetal phase WSM2 for the volumes $V_3$ and $V_4$, the spin-orbit split bands cross now also in the valence band.
As a consequence, from $V_3$ to $V_1$ we have a surface Lifshitz transition in the topological regime. Such a topological Lifshitz transition has already been observed in other Weyl semimetal compounds~\cite{Yang2019,wadge2022topological}.


\begin{figure}{l}
  \vspace{-20pt}
  \begin{center}
        \includegraphics[width=.95\linewidth]{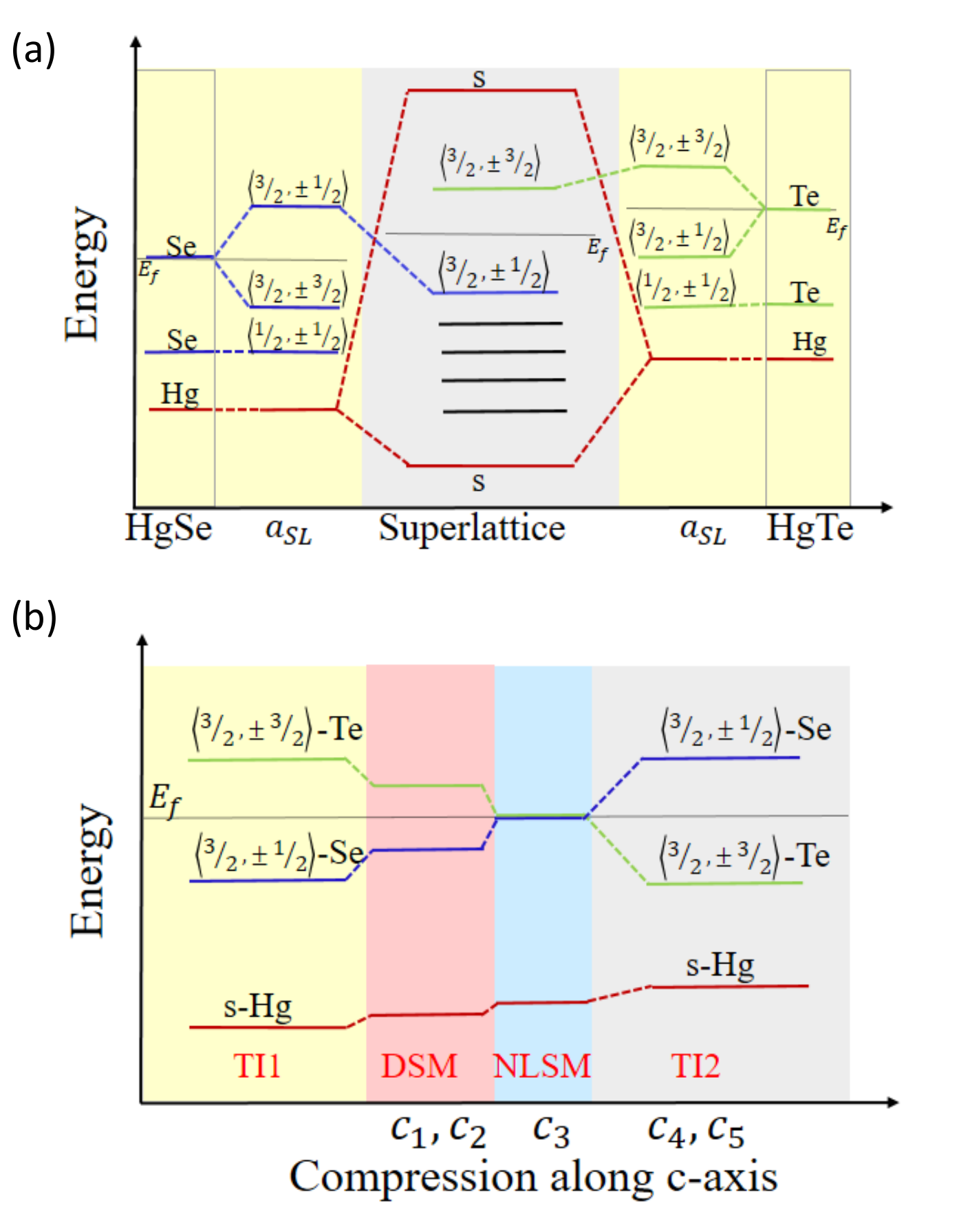}
  \end{center}
   \caption{(a) Schematic of energy levels in the superlattice with additional band inversion: the outer boxes show the energetic levels $\Gamma_6$, $\Gamma_7$ and $\Gamma_8$ in bulk HgTe and HgSe. Strain as an effect of changing the lattice constant of the superlattice (a$_{SL}$) splits the energy levels, while the interface hybridization (center) rearranges them. (b) Schematic of the energetic levels of $s$-Hg, $\ket{3/2,\pm1/2}$-Te and $\ket{3/2,\pm3/2}$-Se states as a function of the $c$ lattice constant. The colors indicate the different topological phases DSM, NLSM and TI. Note that the energy difference between $\ket{3/2,\pm1/2}$-Te and $\ket{3/2,\pm3/2}$-Se is not the crystal field, since the orbitals belong to different chalcogenide atoms. The colors red, green and blue denote the main orbital character of the energetic levels, namely Hg, Te and Se, respectively.}
   \label{energylevels}
\end{figure}

Another difference between the Weyl phases concerns the structure of surface states. The band structures of the slabs for the surface orientations (100) and (001) are reported in ~\Cref{fig : L5.60}(c)-(d), respectively. The corresponding Fermi surfaces for the surface orientation (100) and (001) are shown in ~\Cref{fig : L5.60}(e)-(f), respectively.
~\Cref{fig : L5.60}(g) shows the magnification of the (001) Fermi surface close to $\bar{\Gamma}$.

We observe that, with respect to WSM2 (compare to \Cref{fig : L6.27}), there is a difference in the number of Fermi lines and in the connectivity of the Weyl points.
For the (100) slab, we now observe two additional large circular Fermi rings encompassing the Weyl points and their Fermi arcs, whereas WSM2 showed only Fermi arcs.
The (001) slab has one additional Fermi ring and the connectivity of the Weyl points has changed: instead of two separate pairs of Weyl node projections intra-connected by two Fermi arcs, the Fermi arcs now connect all the Weyl-node projections in a chain-like fashion.
Consequently, for both terminations the number of crossings along a given line through half of the surface BZ is the same as for WSM2 modulo 2.
Hence, the infinite superlattice topology of the two WSM phases WSM1 and WSM2 is indeed the same~\cite{LKBO2017}.
Also the Berry flux reported in ~\Cref{fig : L5.60}(h) is similar to the one for the volumes $V_3$ and $V_4$, confirming that the two WSM phases have the same topological properties.

Shrinking the volume further, we observe the $\Gamma_6$ band coming closer to the Fermi level. For the volume $V_1$, there is still a multiple band inversion between the $p$-orbitals and the $\Gamma_6$ band, as well as a band inversion involving $\ket{3/2,\pm3/2}$-Se and $\ket{3/2,\pm1/2}$-Te orbitals close to the Fermi level. A further reduction of the volume to unrealistic values pushes the $\Gamma_6$ level above the Fermi level creating a trivial insulator. This transition is favored if layers of trivial insulators, such as CdTe, are inserted into the heterostructure.

\begin{figure*}[!ht]
  \centering
  \includegraphics[width=0.95\linewidth]{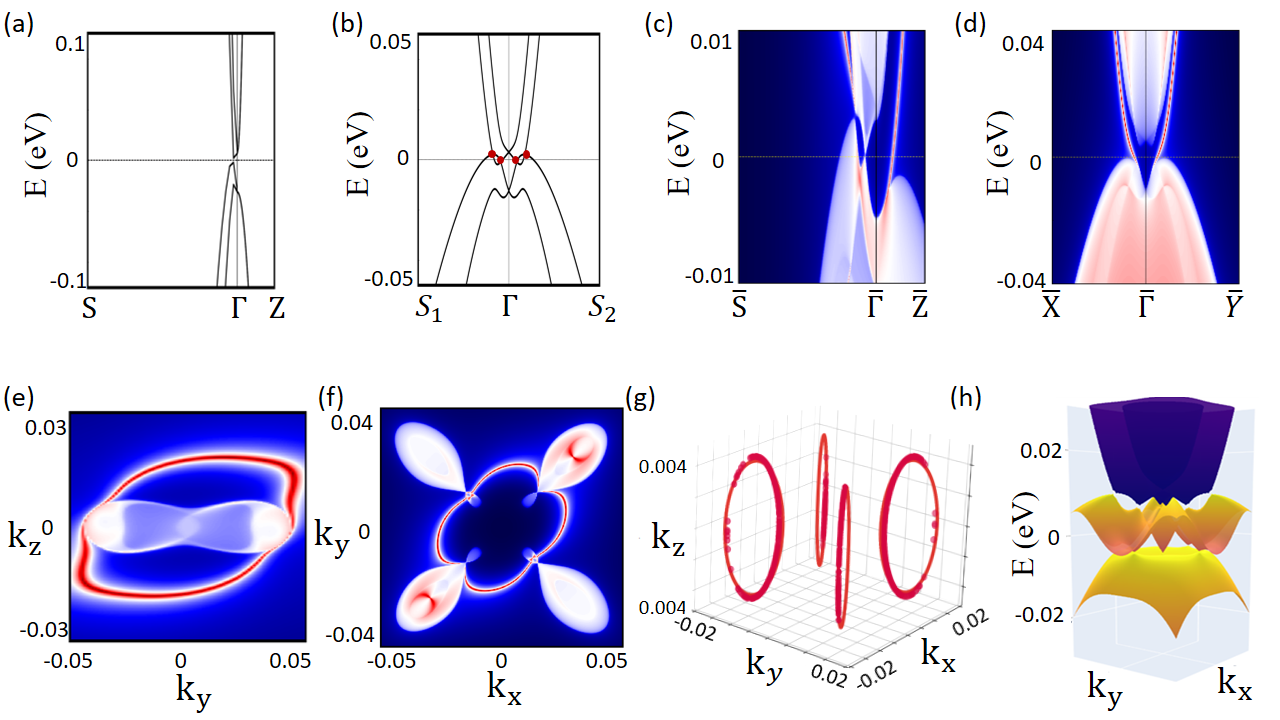}
  \caption{Band structure of the HgTe/HgSe superlattice with 5\% compressive strain (a) along the high-symmetry lines S-$\Gamma$-Z. (b) Wannier band structure in the plane at $k_z=0$ plane where S$_1=(0.085\frac{\pi}{a}, 0.085\frac{\pi}{b},0)$ and S$_2$=$(-0.085\frac{\pi}{a}, -0.085\frac{\pi}{b},0)$. The nodal points are highlighted by red dots. Band structure of slabs with (c) (100)  surfaces and  (d) (001)  surfaces. 
  Fermi surfaces of the slabs with (e) (100) surfaces and with (f) (001) surfaces. (g) Nodal lines in the full BZ. (h) Band structure at the $k_z$ = 0 plane.}
  \label{fig : C5}
\end{figure*} 

\subsection{Topological properties as a function of uniaxial strain}

To illustrate the effect of the strain, we first discuss the order of the energetic levels at the $\Gamma$ point for the volume V$_2$. While the energetic levels can be labelled by their parities in the presence of inversion symmetry ~\cite{PhysRevB.82.045122}, this is not possible here. We start from the bulk HgTe and bulk HgSe energy levels $\Gamma_8$, $\Gamma_7$ and $\Gamma_6$ as shown in the outer parts of the top panel in Fig.~\ref{energylevels}(a). HgSe is more electronegative, so the energetic levels of HgSe are lower. Assuming the electrons can rearrange along the c-axis as it happens in the case of an interface, the effect of the strain on the energy levels $\ket{3/2,\pm3/2}$ and $\ket{3/2,\pm1/2}$ coming from the change of the lattice constant a$_{SL}$ is different for HgTe and HgSe.
Finally, considering also the hybridization between the two components of the superlattice, i.e., HgSe and HgTe, we obtain the final order of the energetic levels, which is shown in the center of Fig.~\ref{energylevels}(a).
In terms of orbital weights, we note that the states with s-orbital character are much more delocalized than the states with p-orbital character in zinc-blende superlattices~\cite{Hussain22}. Moreover, at the $\Gamma$ point, the states with s-orbital character are decoupled from the states with p-orbital character. We further note that the final superlattice states with p-orbital character are mixtures of Te- and Se-orbitals, while Figure~\ref{energylevels}(a) indicates only the main character (Te or Se) of the energetic levels. 

Bulk HgTe realizes a type-I WSM phase at large uniaxial tensile strain and a TI phase at large uniaxial compressive strain, where the strain is with respect to the c-axis.
In between, the material is a type-II WSM\cite{kirtschig2016surface,Ortix2014absence,ruan2016ideal}. Note that our strain notation is different from Ref.~\onlinecite{ruan2016ideal}.

In this subsection, we investigate the evolution of the topological phases in the HgTe/HgSe superlattice as a function of uniaxial strain along the c-axis.
We start from the superlattice at the compressed volume $V_2$. We note, however, that we obtain the same progression of topological phases starting from the unstrained volume $V_3$.
At the volume $V_2$, the heterostructure realizes a small-gap TI, which can be viewed as an approximate Dirac semimetal. 
We have performed calculations for compressed $c$-axes corresponding to strain up to 7\%.
For small strain, we find that the energy gap closes at the Dirac points and the system remains a Dirac semimetal for an extended range of strain values.
Subsequently, the superlattice realizes a nodal-line semimetal phase at around 5\% of compression.
Increasing the compressive uniaxial strain further,
the HgTe/HgSe superlattice evolves into another TI phase with a sizeable energy gap.
The TI phase of the HgTe/HgSe superlattice is similar to the one observed in strained HgTe.
However, the path to arrive at the topological insulator phase is different as we will show below.
A schematic diagram of the different phases is shown in \Cref{energylevels}(b).
In the following, we analyze the arising phases in more detail.


\begin{figure*}[!ht]
  \centering
  \includegraphics[width=0.95\linewidth]{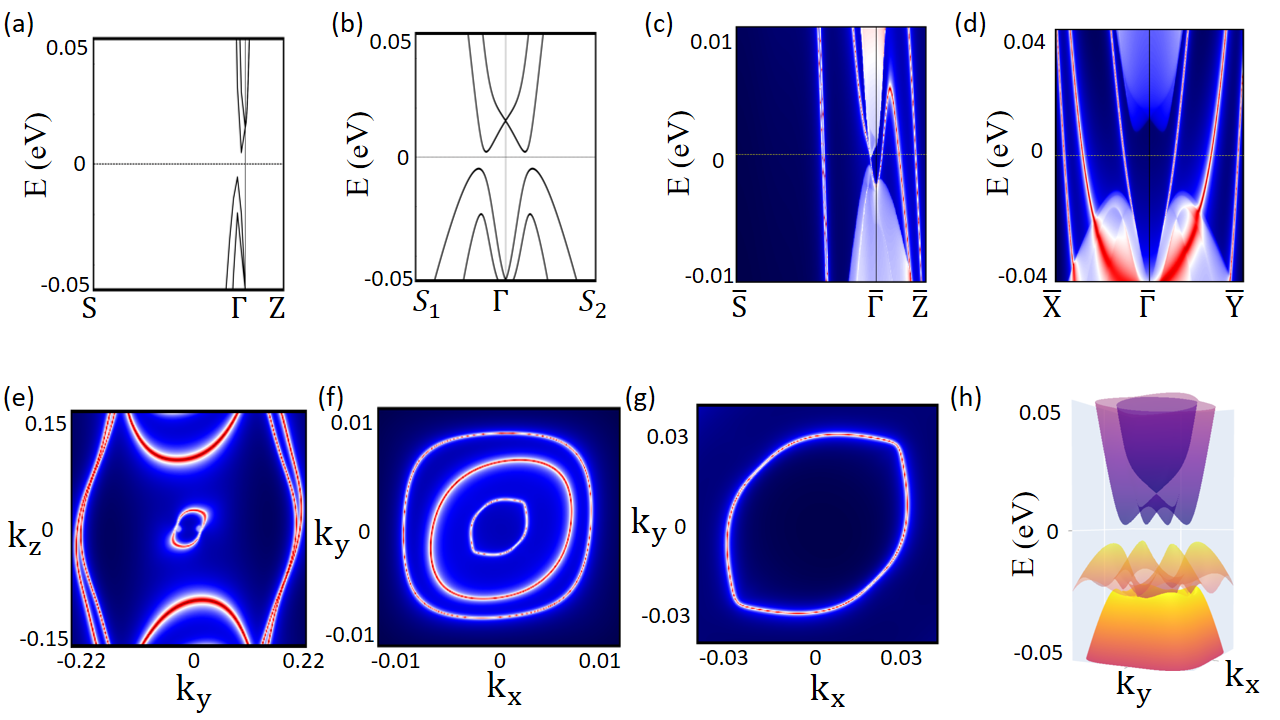}
\caption{Band structure of the HgTe/HgSe superlattice with 7\% compressive strain (a) along the high-symmetry lines S-$\Gamma$-Z. (b) Wannier band structure at $k_z$=0 where S$_1$=$(0.085\frac{\pi}{a}, 0.085\frac{\pi}{b},0)$ and S$_2$=$(-0.085\frac{\pi}{a}, -0.085\frac{\pi}{b},0)$. Band structure of slabs with (c) (100) surface and  (d) (001)  surface. Fermi surfaces of the slabs with (e) (100) surfaces and with (f) (001) surfaces. (g) Magnification of panel (f). (h)  Band structure in the $k_z=0$ plane.}
  \label{fig : C7}
\end{figure*}

\subsubsection{Nodal-line semimetal at 5\% compression}

At 5\% of compression, the minimal gap along the high-symmetry lines moves from the Z-$\Gamma$ direction to the $\Gamma$-S direction as shown in \Cref{fig : C5}(a).
The reason is that, at the $\Gamma$ point, the $\ket{3/2,\pm3/2}$-Se orbital goes below the $\ket{3/2,\pm1/2}$-Te orbital thereby undoing the band inversion between the associated bands.
As a consequence, the bands lose their camel-back shape along the $\Gamma$-Z line to obtain a large Rashba band along the $\Gamma$-S line.
The upper Rashba bands of the valence bands intersect the lower Rashba bands of the conduction band producing the features of the NLSM phase.
Indeed, we find four nodal loops with a linear dispersion rendering this phase a NLSM.
A path cutting through the nodal loops is shown in \Cref{fig : C5}(b), where the nodal points are highlighted by red dots.
The band structure in the $k_z=0$ plane is illustrated in \Cref{fig : C5}(h). The four nodal lines are located in the $k_x=\pm{k_y}$ planes.
A visualization of the four nodal lines in the full BZ is provided in \Cref{fig : C5}(g). These nodal lines are close to the Fermi level but are not isoenergetic.
This is similar to the situation in bulk HgTe under tensile strain along the (111) axis \cite{zaheer2013spin}, 
but the nodal lines in our heterostructure have a different structure.
Furthermore, the nodal lines are protected by the $M_{xy}$ and $M_{x\bar{y}}$ mirror symmetries.  The band structure for slabs with (100) and (001) surface are reported in \Cref{fig : C5}(c,d), respectively. The corresponding Fermi surfaces are shown in \Cref{fig : C5}(e,f).

\subsubsection{Topological insulator at 7\% compression}

At 7\% of compression, the gap along the $\Gamma$-S direction increases as shown in \Cref{fig : C7}(a).
The nodal lines have shrunk to points and gapped out, as can be seen in \Cref{fig : C7}(b) when compared with \Cref{fig : C5}(b).
The superlattice now realizes a topological insulator with a sizable gap and a single band inversion, which we denote TI2. This phase has properties similar to strained bulk HgTe\cite{ruan2016ideal} with a direct band gap of 10 meV and strong
topological invariants ($\nu_0$; $\nu_1\nu_2\nu_3$) = (1;000).
The band structure in the $k_z=0$ plane is shown in \Cref{fig : C7}(h).
The electronic surface states for the (100) and (001) surface orientations are shown in \Cref{fig : C7}(c,d) and the associated Fermi surfaces are displayed in \Cref{fig : C7}(e,f,g).
For the (100) termination, we observe a Fermi ring centered at $\bar{\Gamma}$ associated with a single Dirac cone. Furthermore, there are additional trivial surface states.
The (001) surface shows an odd number of Fermi rings centered at $\bar{\Gamma}$. 
However, only the central ring is associated with a Dirac cone, whereas the others are trivial surface states, as can be seen from comparing with \Cref{fig : C7}(d).

\section{Conclusion and Outlook}
\label{sec:conclusions}

We have investigated the evolution of topological phases in short-period HgTe/CdTe and HgTe/HgSe superlattices. The HgTe/CdTe superlattice realizes a nodal-line semimetal with two isoenergetic, circular nodal lines at the Fermi level. Under compressive hydrostatic pressure, the superlattice becomes a trivial insulator. The HgTe/HgSe superlattice, in contrast, hosts a richer phase diagram and supports a plethora of topological phases with a Rashba-Dresselhaus spin-orbit splitting close to the Fermi level. The unstrained superlattice is an ideal Weyl semimetal with eight symmetry-related Weyl points at the Fermi level. It is a promising candidate for the realization of the 3D quantum Hall effect~\cite{ma2021three}. Under compression through hydrostatic pressure, the system becomes a small-gap topological insulator close to a Dirac phase as a consequence of an accidental reduction of the bulk inversion asymmetry. Further compression of the volume leads to another Weyl semimetal phase with the same bulk topological properties as the unstrained superlattice but with different Fermi-arc connectivities on the surface. Applying compressive uniaxial strain to the Dirac semimetal phase, the heterostructure goes through a nodal-line semimetal phase and a topological insulating phase with a single band inversion.

The nodal-line phases found in this work are relevant to the search for materials with carriers residing in flat energy bands which can support new correlation-driven collective phases~\cite{regnault2021catalogue} which can support new correlation-driven collective phases, as recently proposed by some of the present authors.\cite{lau2021designing}
The nodal-line phases in the HgTe/CdTe superlattices as well as the uniaxially strained HgTe/HgSe superlattices are promising candidates for the realization and study of three-dimensional flat bands, which could give rise to exotic, strongly correlated phases due to an enhancement of electronic interactions. 
Motivated by recent experimental progress in the growth of II-VI strained nanostructures by molecular beam epitaxy~\cite{Plachta:2018_N,Hajer:2019_NL} and high electron mobility systems~\cite{Betthausen:2014_PRB,shamim2020emergent}, the synthesis of core-shell nanowires of the HgTe-based superlattices is a possible way to create the required bend to obtain the flat bands. In fact, nanowires of II-IV semiconductors have recently been fabricated\cite{PhysRevMaterials.1.023401,PhysRevMaterials.4.066001}.
Moreover, the superconductivity induced by the flat bands in a band-inverted nanowire could be a platform to host topological superconductivity and Majorana zero modes. 

As another promising research direction, 2D quantum wells of HgTe/HgSe with a thickness larger than the critical thickness and sandwiched between CdTe could generate a topological insulator with multiple band inversions and host new exotic phases beyond the quantum spin Hall effect.  Experimentally, a challenge will be the fabrication of a sufficiently sharp interface between HgTe and HgSe, since the system will tend to create a digital alloy. Nonetheless, we expect a nearly ideal Weyl semimetal phase, which is topologically protected and appears at zero strain, to be achievable even in the case of a digital alloy.

\section*{Acknowledgments}
We acknowledge T. Wojtowicz, Z. Yu, T. Hyart and C. M. Canali for useful discussions. The work is supported by the Foundation for Polish Science through the International Research Agendas program co-financed by the European Union within the Smart Growth Operational Programme. This work was financially supported by the National Science Center in the framework of the "PRELUDIUM" (Decision No.: DEC-2020/37/N/ST3/02338). G. C. acknowledges financial support from ''Fondazione Angelo Della Riccia''. A. L. acknowledges support from a Marie Sk{\l}odowska-Curie Individual Fellowship under grant MagTopCSL (ID 101029345). W. B. acknowledges the support by Narodowe Centrum Nauk (NCN, National Science Centre, Poland) Project No. 2019/34/E/ST3/00404. 
The work at Northeastern University was supported by the Air Force Office of Scientific Research under award number FA9550-20-1-0322 and benefited from the computational resources of Northeastern University's Advanced Scientific Computation Center (ASCC) and the Discovery Cluster. 
The work at TIFR Mumbai is supported by the Department of Atomic Energy of the Government of India under project number 12-R$\&$D-TFR-5.10-0100.
We acknowledge the access to the computing facilities of the Interdisciplinary Center of Modeling at the University of Warsaw, Grant No.~GB84-1 and No.~GB84-7.
We acknowledge the CINECA award under the ISCRA initiative IsC81 "DISTANCE" Grant, for the availability of high-performance computing resources and support.\\

\textit{Data availability:} The data shown in the figures is available at Ref.~\onlinecite{zenodo}.

\begin{appendices}

\begin{table}
	\centering
	\begin{tabular}{|c|c|c|c|c|c|}
		\hline		  $E_{XC}$ & Material & $E_g$ (meV) & $\Delta_{SOC}$ (meV) & a (\AA)\\
		 \hline
		 \multirow{4}{*}{GGA}
		  & HgTe & -976  & 743 & 6.46 \\
		  & HgSe & -1028 & 219 & 6.08 \\
		  & HgS &  -429  & 117 & 5.85 \\
		  & CdTe & +1119 & 821 & 6.60 \\
		 \hline
		 \multirow{4}{*}{MBJGGA}           & HgTe & -315 & 708 & 6.46 \\
		 & HgSe & -235 & 183 & 6.08 \\
		 & HgS  & +429 & 290 & 5.85 \\	   & CdTe & +1267 & 801 & 6.60 \\
		 \hline
	\end{tabular}
\caption{Values of band inversion strength $E_g$, spin-orbit constant $\Delta_{SOC}$ and lattice constant $a$ of bulk zinc blende structures for two different exchange-correlation functionals (GGA and MBJGGA).}
\label{Tab:Table_lattice_constant}
\end{table}

\section{Computational details}
Electronic structure calculations were performed within the framework of the first-principles density functional theory based on a plane wave basis set and the projector augmented wave method using VASP\cite{VASP} package. The calculation is fully relativistic by considering spin-orbit coupling (SOC). A plane-wave energy cut-off of 250~eV has been used. As an exchange-correlation functional, the generalised gradient approximation (GGA) of Perdrew, Burke, and Ernzerhof has been adopted\cite{perdew1996generalized}. We have performed the calculations using 6$\times$6$\times$4 $k$-points centered at $\Gamma$ with 144 $k$-points in the BZ for the superlattice with 24 atoms shown in ~\Cref{heterostruc}(a).


The band order and band gap of the bulk have been validated with a meta-GGA approach, which is the modified Becke-Johnson exchange potential together
with local density approximation for the correlation potential scheme\cite{MBJLDA}. Indeed, we have performed the calculation using MBJGGA with the parameter $C_{MBJ}=1.11$ in order to get the experimental band ordering for both HgTe and HgSe.
The GGA band ordering of HgSe is  $\Gamma_8,\Gamma_7,\Gamma_6$ in agreement with the experimental band ordering. However, the GGA band ordering of the HgTe is $\Gamma_8$, $\Gamma_7$ and $\Gamma_6$ while the experimental band ordering is $\Gamma_8$, $\Gamma_6$ and $\Gamma_7$. The electronic properties of the HgTe/HgSe superlattice are not affected by the use of the GGA since the bands near the Fermi level of the superlattice are dominated by the bands associated with $\Gamma_8$ which  remains unchanged in both functionals. \\

The electronic structure calculation of the heterostructure was performed using both GGA and meta-GGA for the volume $V_3$ discussed in the main text. Within GGA and meta-GGA, we obtain the same qualitative results for the volume $V_3$. Therefore, we considered the GGA exchange-correlation functional throughout this paper. We study the systems without structural relaxation not to add further degrees of freedom and complexity to the topological phase diagram.

We extracted the real space tight-binding Hamiltonian with atom-centred Wannier functions with $s$-like cation and $p$-like anion orbital projections using the VASP2WANNIER90 code\cite{mostofi2008wannier90}. The topological properties are studied using the Wanniertools package \cite{wu2018wanniertools}. The surface states are obtained within the iterative Green's function approach\cite{Greenwanniertools}. We used a denser $k$-point grid 16$\times$16$\times$4 to generate the model Hamiltonian with the Wannier basis. Since both the (100) and the (001) surface are polar, the anion and cation terminations differ from a quantitative point of view including small differences in the Fermi level and different number of surface states. In the main text, we present results for the terminations that show the best visualizations for our purposes.

\section{Topological bulk properties}

The bulk properties were analyzed using the experimental lattice constants of HgX (i.e., X = Te, Se, S) materials with zinc blende crystal structure. We have calculated the band gap $(E_{g})$, the band order and the spin-orbit coupling $(\Delta_{SOC})$. The values are tabulated in \Cref{Tab:Table_lattice_constant}. The band inversion between Hg-s orbital and X-p orbital occurs at the $\Gamma$ point which indicates a topological insulator $(TI)$ phase in this class of materials. According to the symmetry of the wave function, the Hg-$s$ or Cd-$s$ bands are labeled as $\Gamma_{6}$ and the Te-$p$, Se-$p$ or S-$p$ bands split into $\Gamma_{8}(j=\frac{3}{2})$ and $\Gamma_{7}(\frac{1}{2})$ bands with SOC.
We define the following quantities
\begin{equation}
\Delta_{SOC} =\Gamma_{8}-\Gamma_{7}
\end{equation}
\begin{equation}
		E_{g}=\Gamma_{6}-\Gamma_{8}
\end{equation}
\vspace{.2cm}
which we report in \cref{Tab:Table_lattice_constant}.

\begin{figure}[!t]
  \centering
  \includegraphics[width=0.95\linewidth]{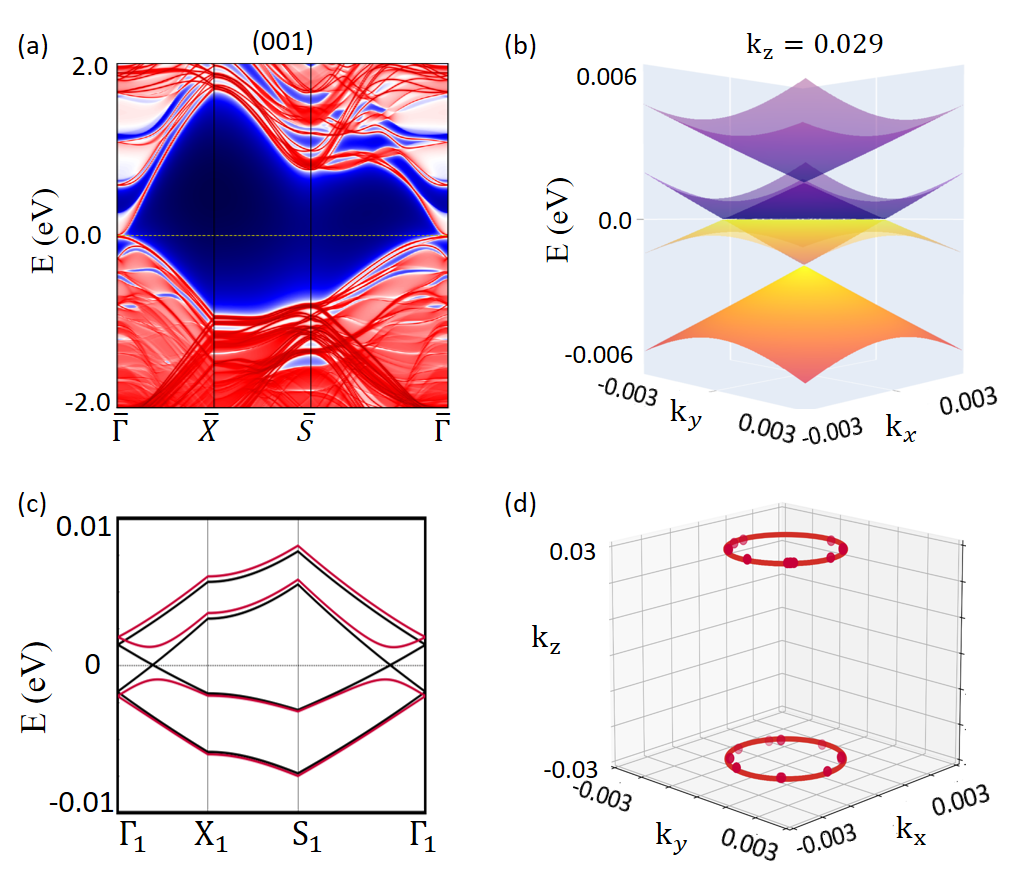}
  \caption{(a) Surface band structure of the (001) Te-terminated surface of the (HgTe)$_4$/(CdTe)$_4$ superlattice. (b) Band structure of the infinite superlattice in the $E-k_x-k_y$ space at fixed $k_z$=0.029 \AA$^{-1}$. (c) Band structure with nodal points at $k_z$=0.029 \AA$^{-1}$(black) and gapped at $k_z$=0.030 \AA$^{-1}$(red). These four bands are a set of bands isolated from the rest of the band structure. (d) Nodal lines of the infinite superlattice at the $k_z$=$\pm$0.029 \AA$^{-1}$ planes.}
  \label{fig : 4l_CdTe_L6.60}
\end{figure}

In the case of a trivial insulator, the cation $s$-type of bands lies  above the anion $p$-type of bands while the scenario is opposite for the topological insulating phase.
HgTe and HgSe are topological zero-gap semimetals with positive spin-orbit gap.
However, HgS has negative spin-orbit gap and it is a trivial insulator in agreement with experimental results.

The volume can tune the effective spin-orbit gap via the hopping between d-electrons of Hg and p-electrons of the chalchogen atoms.
The bare SOC of S is positive, but becomes negative due to the
$p$-$d$ hopping.
However, the spin-orbit couplings of Te and Se are one order of magnitude larger than the variation that can be induced by these effects. Therefore, these are not relevant effects in this case.

Another way to tune the effective spin-orbit coupling is uniaxial pressure, which changes the crystal field energy difference $\varepsilon_{x,y}$-$\varepsilon_z$, but also in this case the effect is too small to change significantly the SOC of Te and Se.

\begin{figure*}[!ht]
  \centering
  \includegraphics[width=0.95\linewidth]{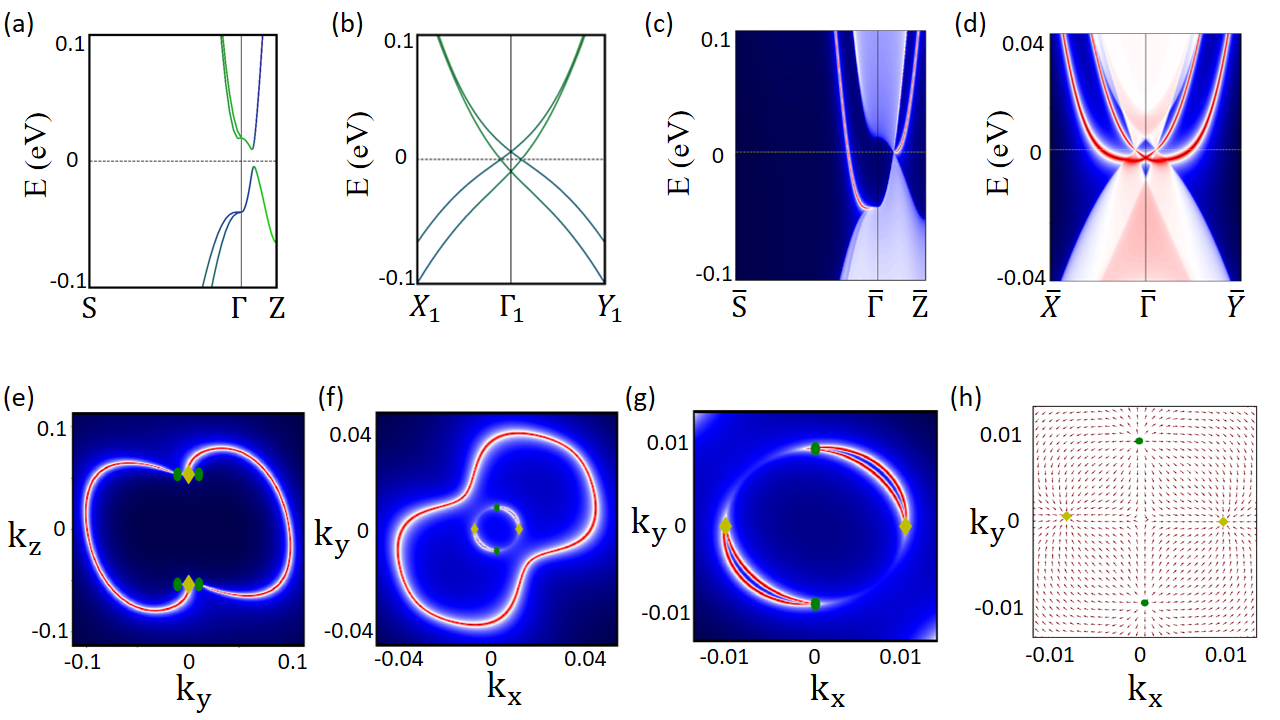}
  \caption{Band structure of the HgTe/HgSe infinite superlattice at the volume $V_4$ (a) along the  S-$\Gamma$-Z high-symmetry lines. b) Band structure in the plane of the Weyl points at $k_z=k_z^*=0.0536$ {\AA$^{-1}$}.
  Band structure projected onto the c) (100) surface and d) (001)  surface orientation. Fermi surface of the slab for the e) (100) surface f) of the (001) surface orientation. Red means presence of electronic states while blue means absence of electronic states. g) Magnification of the previous panel where we can clearly see the Weyl points and their connectivity. h) In-plane component of the Berry flux at the $k_z=k_z^*$ plane including four Weyl points. The Weyl points located at $(\pm {{k_\parallel}^*},0,\pm {k_z^*})$ have chirality -1 (yellow diamond marker), while the Weyl points located at $(0,\pm {{k_\parallel}^*},\pm {k_z^*})$ have chirality +1 (green circle marker).}
  \label{fig : L6.60}
\end{figure*}

\begin{figure*}[!ht]
  \centering
  \includegraphics[width=0.95\linewidth]{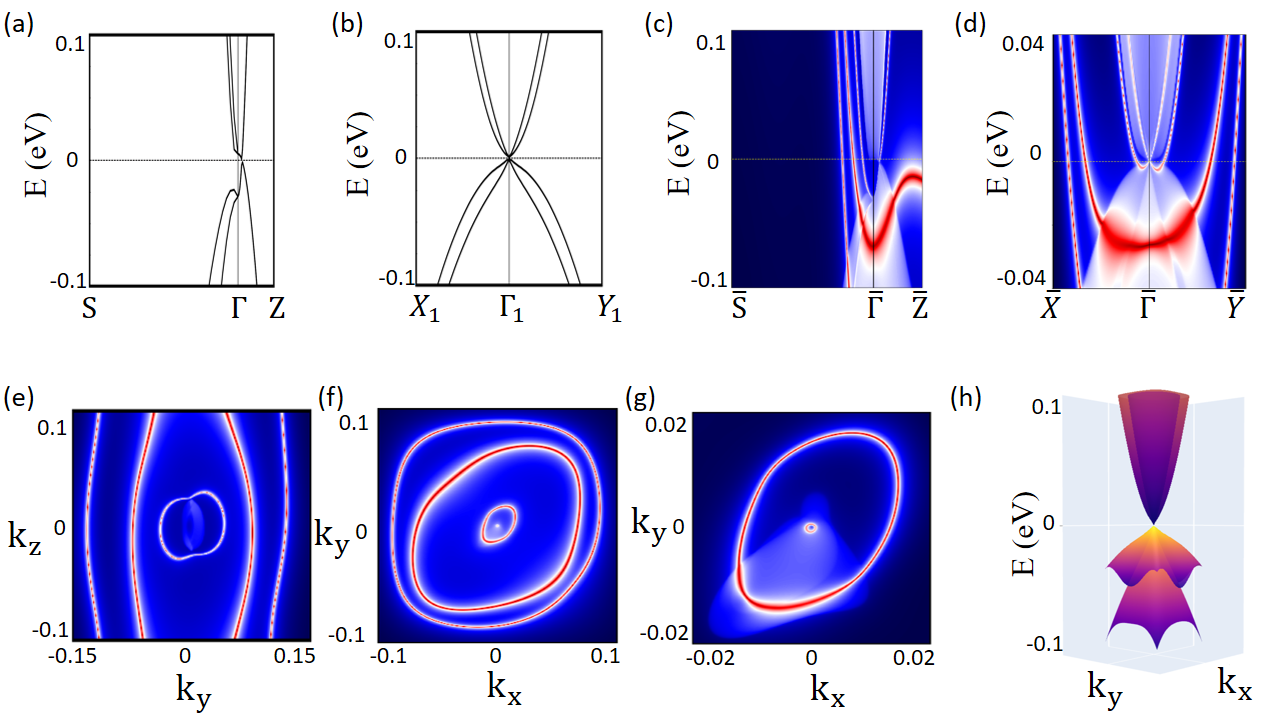}
  \caption{Band structure of the HgTe/HgSe superlattice with 3\% compressive strain (a) along the high-symmetry lines S-$\Gamma$-Z and (b) in the plane of the Dirac points at $k_z$=k$_z^*$= 0.021 \AA$^{-1}$. Band structure for slabs with (c) (100) surfaces and (d) (001)  surface. Fermi surfaces of the slabs with (e) (100) surfaces and with (f) (001)  surfaces. (g) Magnification of panel (f). (h) Band structure in the $k_z^*$=0.021 \AA$^{-1}$ plane.}
  \label{fig : C3}
\end{figure*}

\begin{figure*}[!ht]
  \centering
  \includegraphics[width=0.95\linewidth]{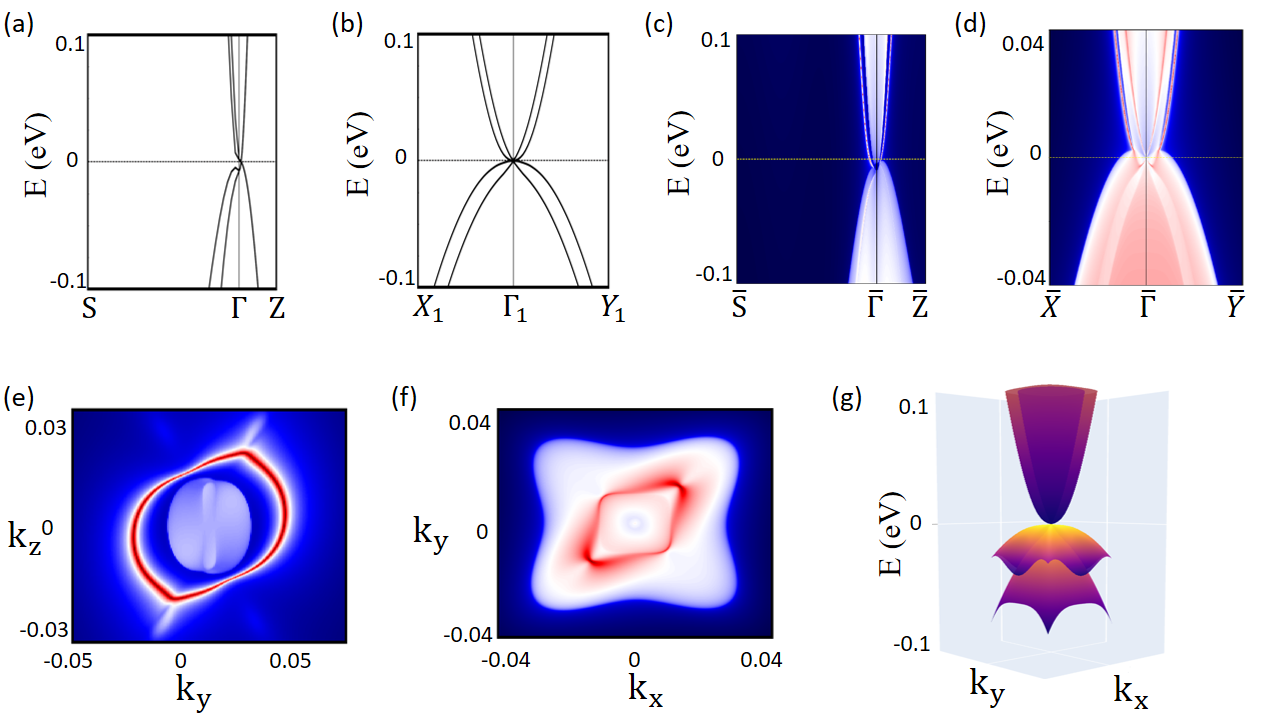}
  \caption{Band structure of HgTe/HgSe superlattice with 4\% compressive strain (a) along the high-symmetry lines S-$\Gamma$-Z. (b) and in the plane of the Dirac points at $k_z=k_z^*= 0.010$~\AA$^{-1}$. Projected surface bands onto the (c) (100) surface and (d) (001) surface orientation. Fermi surface of the slab for the (e) (100) surface (f) of the (001) surface orientation. (g) Band structure at the $k_z=k_z^*=0.010$~\AA$^{-1}$ plane.}
  \label{fig : C4}
\end{figure*}

\section{Nodal line for (H\lowercase{g}T\lowercase{e})$_4$/(C\lowercase{d}T\lowercase{e})$_4$ supercell}

Here we discuss the isoenergetic nodal line present in the (HgTe)$_4$/(CdTe)$_4$ superlattice. The results are qualitatively similar to the (HgTe)$_3$/(CdTe)$_3$ results provided in Fig.~\ref{fig : CdTe_L6.60_surface} of the main text. The band structure projected on the (001) surface shows topological surface states connecting valence and conduction bands as reported in Fig.~\ref{fig : 4l_CdTe_L6.60}(a).
The band structure has two isoenergetic circular nodal-lines with the same radius in the planes k$_z^*$=$\pm$0.029 {\AA$^{-1}$} as shown in Fig.~\ref{fig : 4l_CdTe_L6.60}(b) at fixed k$_z$ and in Fig.~\ref{fig : 4l_CdTe_L6.60}(c) along specific lines through the Brillouin zone. The nodal-line radius is smaller than in the case of (HgTe)$3$/(CdTe)$_3$, as can be seen in Fig.~\ref{fig : 4l_CdTe_L6.60}(d). Therefore, we expect it to further shrink with increasing number of layers until it disappears.

\section{Weyl semimetal phase in H\lowercase{g}T\lowercase{e}/H\lowercase{g}S\lowercase{e} at the volume $V_4$}

\Cref{fig : L6.60}(a) shows the band structure for the volume $V_4$ along the high-symmetry lines S-$\Gamma$-Z.
Considering just the high-symmetry lines in the \text{k}-space, the minimum bandgap is
13.2 meV along the $\Gamma$Z direction. A linear dispersion  with gap closing typical of Weyl points is found along the $X_1$-$\Gamma_{1}$-$Y_1$ direction as shown in ~\Cref{fig : L6.60}(b), where $\Gamma_{1}$=(0,0,$k_z^*$), X$_1=(0.2\pi/a,0,k_z^*)$ and Y$_1=(0,0.2\pi/a,k_z^*)$. For the volume $V_4$, we found that $k_z^*=0.0536$ {\AA$^{-1}$}. The crossing point between the bands at the Fermi level determines the position of the Weyl points. 

The presence of a topologically protected Fermi arc is a hallmark of the Weyl semimetal phase. To confirm the topological nature of the 3D superlattice, we have calculated the surface electronic states and Fermi arcs for the (100) and (001) surface orientations considering the notation of the conventional unit cell as shown in \Cref{heterostruc}(a).
The (010) surface is equivalent to the (100) surface.
The band structure of the slab for the surface orientation (100) with chalcogenides termination is shown in \Cref{fig : L6.60}(c).  The topological surface states connecting the valence and the conduction bands confirm the topological nature of the system, moreover we have a gapless point at the coordinates $(0,k_z^*)$ projected on the 2D BZ of the slab.
The Fermi surface for the slab with surface orientation (100) is shown in \Cref{fig : L6.60}(e). It has six gapless points, four points with projected coordinates $(\pm {k_\parallel}^*,\pm k_z^*)$ and monopole charge +1 and two points with coordinates $(0,\pm k_z^*)$ and monopole charge -2, where ${k_\parallel}^*=$0.0105 {\AA$^{-1}$}. Open Fermi arcs are observed. A clear connectivity between the monopole charge -2 and
one of the monopole charges +1 is visible. However, the other connectivity is not clearly visible due to the short distance between the Weyl points for this surface orientation.

The band structure for the slab with (001) surface orientation is shown in \Cref{fig : L6.60}(d). Two gapless points are present at $({k_\parallel}^*,0)$ and $(0, {k_\parallel}^*)$.
The Fermi surface of the slab with orientation (001) is shown in \Cref{fig : L6.60}(f) and its magnification is in \Cref{fig : L6.60}(g).
We can see one closed Fermi surface due to the bulk topology and the Weyl points around $\bar{\Gamma}$.
The Weyl points with monopole charge +2 are projected at $(0,\pm{{k_\parallel}^*})$, while the Weyl points with monopole charge of -2 are projected at $(\pm{{k_\parallel}^*},0)$ with a clear connection between them.

\begin{figure*}[!ht]
  \centering
  \includegraphics[width=0.95\linewidth]{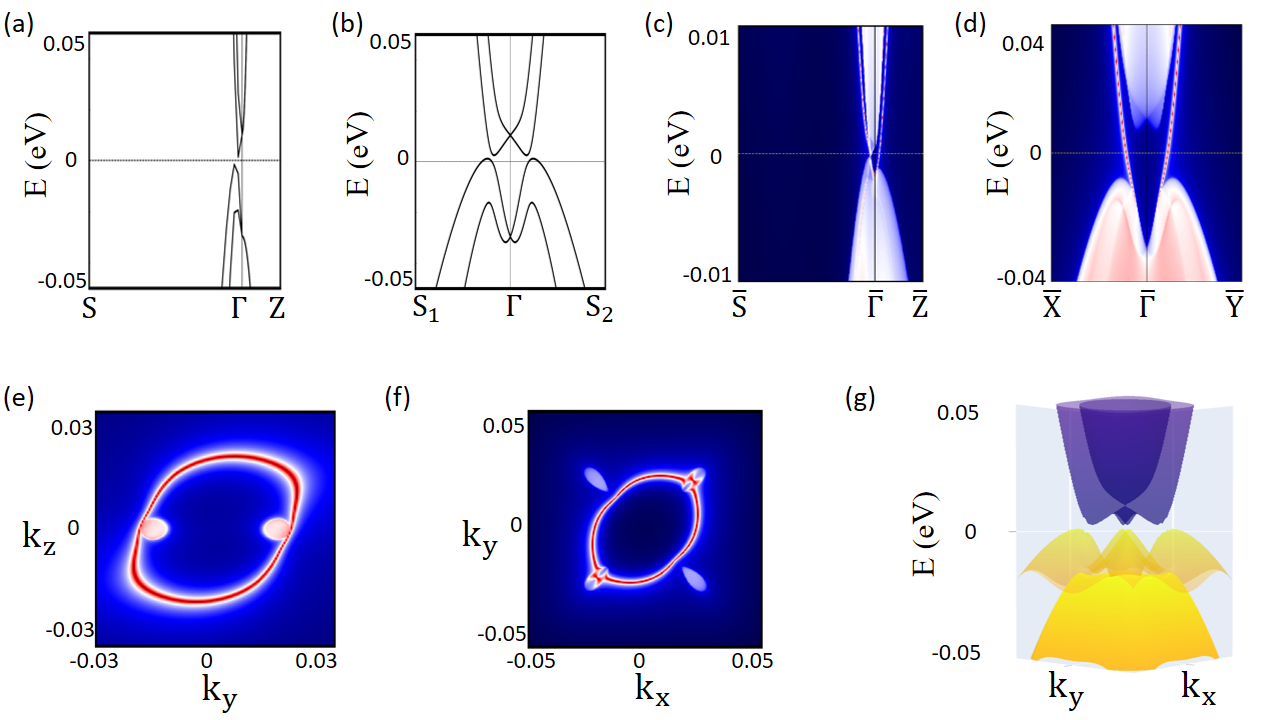}
  \caption{Band structure of HgTe/HgSe superlattice with 6\% compressive strain (a) along the high-symmetry lines S-$\Gamma$-Z. (b) Wannier band structure at $k_z$=0. Projected surface bands onto the (c) (100)  surface and (d) (001)  surface orientation. Fermi surface of the slab for the (e) (100) surface (f) of the (001)  surface orientation. (g)  band structure at the $k_z$=0 plane.}
  \label{fig : C6}
\end{figure*}

Furthermore, we have calculated the Berry curvature in the $k_x-k_y$ plane, at fixed $k_z^*=0.0536$ {\AA$^{-1}$} around the Weyl points shown in ~\Cref{fig : L6.60}(h). Looking at the Berry curvature, the Weyl points with chirality $+1$ behave as a source, while Weyl points with chirality $-1$ as a sink. Due to the absence of trivial Fermi surfaces, the Weyl phase in multilayer 3D superlattice could be experimentally detected. It is also noticeable that the Weyl points are well separated by an in-plane distance between opposite chiralities of 2.2\% of the reciprocal lattice constant.

\section{Dirac semimetal of H\lowercase{g}T\lowercase{e}/H\lowercase{g}S\lowercase{e} at 3\% compression}

Applying compressive strain along the c-axis starting from superlattice volume $V_2$, the phase is gapless at 3\% of strain.
We find a Dirac-like dispersion with a four-fold degeneracy as we can see in \Cref{fig : C3}(a).
The linear dispersion is clearly visible along the path X$_1$-$\Gamma_{1}$-Y$_1$ as shown in ~\Cref{fig : C3}(b), where we have defined $\Gamma_{1}$=(0,0,$k_z^*$), X$_1=(0.1\pi/a,0,k_z^*)$ and Y$_1=(0,0.1\pi/a,k_z^*)$ with $k_z^*= 0.021$ \AA$^{-1}$.
The two Dirac points are located at $(0,0,\pm k_z^*)$.

The electronic states of the slab with (100) surface orientation are shown in \Cref{fig : C3}(c), while the results for the (001) surface orientation are presented in \Cref{fig : C3}(d).
The respective Fermi surfaces are shown in \Cref{fig : C3}(e,f).
For the (100) termination, as compared to the volume $V_2$ (see \Cref{fig : L5.82}), we observe that the Fermi circle associated with the surface Dirac cone is now interrupted by the surface projections of the infinite superlattice Dirac points at $(0,\pm k_z^*)$.
This leads to closed Fermi-arc features characteristic of Dirac semimetals.
For the (001) termination, we find three Fermi rings around $\bar{\Gamma}$, which is similar to what we see for the volume $V_2$.
Notably, the innermost Fermi ring now has a larger radius revealing the projections of the Dirac points at $\bar{\Gamma}$.
In \Cref{fig : C3}(g), the band structure at $k_z= k_z^*$ shows the Dirac point.

\section{Semimetal with parabolic dispersion of H\lowercase{g}T\lowercase{e}/H\lowercase{g}S\lowercase{e} at 4\% compression}

At 4\% compressive strain, the gap along $\Gamma$-Z direction vanishes and our system becomes metallic as we can see in \Cref{fig : C4}(a). The band structure with parabolic behavior along X$_1$-$\Gamma$-Y$_1$ in \Cref{fig : C4}(b) confirms the metallicity. \Cref{fig : C4}(c) and \Cref{fig : C4}(d) show the band structure projected in the (100) and (001) surface orientation, respectively. The associated Fermi surface are represented in \Cref{fig : C4}(e) and (f).  \Cref{fig : C4}(g) represents a magnification of the (001) Fermi surface around the $\Gamma$ point. The 3D band structure with parabolic dispersion is shown in \Cref{fig : C4}(h) for $k_z=k_z^*$=0.010 {\AA}$^{-1}$.

The system shows a coexistence of topological surface states and metallic states. At this lattice constant, the system is turning from the DSM phase to a NLSM phase.

\section{Topological insulator phase of H\lowercase{g}T\lowercase{e}/H\lowercase{g}S\lowercase{e} at 6\% compression}

At 6\% of compression, the band inversion between $\ket{3/2,\pm3/2}$-Se and $\ket{3/2,\pm1/2}$-Te has disappeared as well as the nodal line.
We can still observe a strong Rashba effect in the valence band which is now at the Fermi level, the Rashba in the valence band is unusual but was observed in the literature \cite{autieri2019persistent}.
Therefore, we are left with an insulator with a single band inversion between the s-orbital and the p-orbital as in bulk HgTe with a band gap of 3 meV.
As a consequence, the band crossing is lifted along $\Gamma$S as shown in \Cref{fig : C6}(a,b): the superlattice becomes a TI.
This TI phase is equivalent to the TI phase discussed in the context of strained HgTe \cite{ruan2016ideal}.
The surface states of the (100) and (001) terminations are shown in \Cref{fig : C6}(c,d), and the associated Fermi surfaces are shown in \Cref{fig : C6}(e,f). The 3D band structure is shown in \Cref{fig : C6}(g).

\end{appendices}

\medskip
\newpage
\bibliography{HgTe_Carmine_td}

\end{document}